\documentclass[twocolumn,floatfix,superscriptaddress,a4paper,showpacs,showkeys,nofootinbib,reprint,prc]{revtex4-1}

\usepackage{epsfig}
\usepackage{latexsym}
\usepackage{xspace}
\usepackage[colorlinks=true,linktocpage=true,linkcolor=blue,citecolor=blue,allcolors=blue]{hyperref}
\usepackage[utf8]{inputenc}
\usepackage{indentfirst}
\usepackage{enumerate}
\usepackage{color}
\usepackage{tabularx}

\usepackage{setspace}
\usepackage{lipsum}

\usepackage{amsmath}
\usepackage{amssymb}
\usepackage[english]{babel}
\usepackage{url}
\newcommand{\eq}[1]{\begin{align} #1 \end{align}}

\begin{document}


\title{
Canonical statistical model analysis of p--p, p--Pb, and Pb--Pb collisions at the LHC
}

\author{Volodymyr Vovchenko}
\affiliation{
Institut f\"ur Theoretische Physik,
Goethe Universit\"at Frankfurt, Max-von-Laue-Str. 1, D-60438 Frankfurt am Main, Germany}
\affiliation{Frankfurt Institute for Advanced Studies, Giersch Science Center, Goethe Universit\"at Frankfurt, Ruth-Moufang-Str. 1, D-60438 Frankfurt am Main, Germany}

\author{Benjamin D\"onigus}
\affiliation{
Institut f\"ur Kernphysik,
Goethe Universit\"at Frankfurt, Max-von-Laue-Str. 1, D-60438 Frankfurt am Main, Germany}

\author{Horst Stoecker}
\affiliation{
Institut f\"ur Theoretische Physik,
Goethe Universit\"at Frankfurt, Max-von-Laue-Str. 1, D-60438 Frankfurt am Main, Germany}
\affiliation{Frankfurt Institute for Advanced Studies, Giersch Science Center, Goethe Universit\"at Frankfurt, Ruth-Moufang-Str. 1, D-60438 Frankfurt am Main, Germany}
\affiliation{
GSI Helmholtzzentrum f\"ur Schwerionenforschung GmbH, Planckstr. 1, D-64291 Darmstadt, Germany}

\begin{abstract}
The system-size dependence of hadrochemistry at vanishing baryon density is considered within the canonical statistical model~(CSM) with local exact conservation of three conserved charges, allowing for a possibility of strangeness undersaturation, i.e. $\gamma_S \leq 1$.
Exact baryon number conservation is found to be even more important than that of strangeness in the canonical suppression picture at the LHC, in contrast to intermediate and low collision energies.
The model is applied to p--p, p--Pb, and Pb--Pb data of the ALICE collaboration.
A chemical equilibrium CSM with a fixed $T_{\rm ch} = 155$~MeV describes the trends seen in most yield ratios. However, this vanilla version of CSM predicts an enhancement of the $\phi/\pi$ ratio at smaller multiplicities, in stark contrast to the suppression seen in the data.
The data are described with a 15\% relative accuracy level whence a multiplicity dependence of both the temperature and the strangeness saturation parameter $\gamma_S \leq 1$ is accepted.
Both the canonical suppression and the strangeness undersaturation effects are small at $d N_{\rm ch} / d\eta \gtrsim 100$, but they do
improve substantially the description of hadron yields in p--p collisions, in particular the $\Omega$ yields. 
A possibility to constrain the rapidity correlation volume using net-proton fluctuation measurements is pointed out.
\end{abstract}




\maketitle


\section{Introduction}

A rich body of experimental data on the production of light flavour hadrons produced at LHC energies have recently become available, through the analysis of p--p~\cite{Acharya:2018orn}, p--Pb~\cite{Abelev:2013haa,Adam:2015vsf,Adam:2016bpr}, and Pb--Pb~\cite{Abelev:2013vea,Abelev:2013xaa,ABELEV:2013zaa,Abelev:2014uua} collisions.
This comprehensive set of multiplicity-dependent data does allow for a detailed test of production models. 
Such data are often analysed in the framework of Monte Carlo event generators for p--p collisions, such as Pythia8~\cite{Sjostrand:2007gs,Skands:2014pea}, DIPSY~\cite{Bierlich:2014xba}, or EPOS LHC~\cite{Pierog:2013ria}.
Hadron yield data from central Pb--Pb collisions are
often described on a 10-15\% overall level in statistical models which employ the grand-canonical statistical ensemble ~\cite{Petran:2013lja,Stachel:2013zma,Floris:2014pta}.

The centrality dependence of the chemical freeze-out temperature is usually neglected in the grand-canonical statistical approach, therefore the same constant hadron yield ratios are predicted for all multiplicities.
This evidently cannot describe the observed data, in particular the enhanced production of strange hadrons at higher multiplicities~\cite{ALICE:2017jyt}.
Multiplicity dependence does, however, emerge from a statistical model when exact conservation of conserved charges is treated within the canonical ensemble: canonical corrections to hadron yields become important for sufficiently small reaction volumes,
as was pointed out long time ago~\cite{Rafelski:1980gk,Hagedorn:1984uy}. 
Strangeness enhancement observed in central Pb-Pb collisions has been interpreted as the absence of the canonical suppression effects in large systems~\cite{Hamieh:2000tk}.

The strangeness-canonical ensemble picture has already been applied at the LHC in an ALICE publication~\cite{Acharya:2018orn}, where a qualitative description of the multiplicity dependence of ratios of strange hadron yields to pions was obtained.
In our previous work~\cite{Vovchenko:2018fiy}, the Canonical Statistical Model (CSM), which treats the exact conservation of all three conserved charges, baryon number, electric charge, and strangeness, was applied to the multiplicity dependence of the yields of light (anti-)(hyper-)nuclei. A good qualitative description of the available data was obtained.
The present work extends these two analyses to cover all (stable) light flavoured hadrons measured by the ALICE collaboration in p--p collisions at 7~TeV, p--Pb collisions at 5.02~TeV, and Pb--Pb collisions at 2.76~TeV.
The importance of baryon number conservation for LHC energies is emphasized, in addition to the conservation of strangeness.
Separately, the effects of a multiplicity-dependent chemical freeze-out temperature as well as incomplete chemical equilibration are considered here.

\section{Canonical statistical model}

The standard statistical approach considers an ideal hadron resonance gas (HRG) in thermal and chemical equilibrium at the chemical freeze-out stage.
In the canonical ensemble, the three abelian charges considered -- the baryon number $B$, the electric charge $Q$, and the strangeness $S$ -- are fixed to particular values which are conserved exactly across the so-called correlation volume $V_c$.
The CSM partition function at a given temperature $T$ and correlation volume $V_c$ reads~\cite{Becattini:1995if,Becattini:1997rv}
\eq{\label{eq:Z}
\mathcal{Z}(B,Q,S) & =
\int \limits_{-\pi}^{\pi}
  \frac{d \phi_B}{2\pi}
 \int \limits_{-\pi}^{\pi}
  \frac{d \phi_Q}{2\pi}
  \int \limits_{-\pi}^{\pi}
  \frac{d \phi_S}{2\pi}~
  e^{-i \, (B \phi_B + Q \phi_Q + S \phi_S)} \nonumber \\
  & \quad \times \exp\left[\sum_{j} \sum_{n=1}^{\infty} z_j^n \, e^{i \, n \, (B_j \phi_B + Q_j \phi_Q + S_j \phi_S)}\right].
}
Here the first sum, denoted by index $j$, is over all species included in the list while the second sum over $n$ takes into account the quantum statistics. 
$B_j$, $Q_j$, and $S_j$ are, respectively, the baryon number, electric charge, and strangeness, carried by the particle species $j$, and $z_j^n$ corresponds to the single-particle partition function
\eq{\label{eq:zj}
z_j^n = (\mp 1)^{n-1} \, V_c \, \int dm \, \rho_j(m) \, d_j \frac{m^2 T}{2\pi^2 n^2} \, K_2(n \, m/T).
}
Here $d_j$ is the degeneracy factor for particle species $j$, the minus sign is for fermions and the plus sign is for bosons. 
The integration over the mass distribution $\rho_j(m)$ in Eq.~\eqref{eq:zj} takes into account the finite widths of the resonances.
In the present work we adopt the energy-dependent Breit-Wigner scheme, which was recently advocated for the statistical model description at LHC energies~\cite{Vovchenko:2018fmh}.
The mean multiplicities of various particle species are calculated by introducing fictitious fugacities into the partition function~\eqref{eq:Z} and calculating the corresponding derivatives with respect to these fugacities~(for details see Refs.~\cite{Becattini:1995if,Becattini:1997rv}).
The result is
\eq{
\langle N_j^{\rm prim} \rangle^{\rm ce} = \sum_{n=1}^{\infty} \, \frac{Z(B-nB_j,Q-nQ_j,S-nS_j)}{Z(B,Q,S)} \, n \, z_j^n~.
}
Here the first factors are the canonical chemical factors, which appear due to the requirement of exact conservation of the conserved charges.
The final particle yields, $\langle N_j^{\rm tot} \rangle^{\rm ce}$, are then calculated by including the various feeddown yields stemming from the strong and electromagnetic decays of unstable resonances~(for details see Ref.~\cite{Vovchenko:2018fmh}).
Here we incorporate all hadrons and resonances which have an established status in the 2014 edition of the Particle Data Tables~\cite{Agashe:2014kda}.

The quantum statistical effects at the LHC are quite small for most particles except for the primary pions.
Therefore, here we neglect the quantum statistical effects for baryons, i.e. the sum over $n$ is truncated at the first term for all baryon species.
For the mesons we include 10 terms in the sums for pions, 5 terms for mesons lighter than 1 GeV, and 3 terms for all heavier mesons.
Charmed mesons are not included here.
An implementation of excluded-volume corrections is non-trivial in the canonical ensemble and is not considered in the present work.

One may consider a selective canonical treatment of certain conserved charges, while preserving the grand-canonical treatment of other charges.
For example, to preserve the grand-canonical treatment of the electric charge conservation, let $\phi_Q = 0$ in the integrand in Eq.~\eqref{eq:Z} and introduce the electric charge fugacity factor $\lambda_Q^{n}$ into Eq.~\eqref{eq:zj}.
A selective canonical treatment of conserved charges can clarify the importance of various conservation laws for a given system.
At the LHC the selective canonical treatment has previously been considered only for strangeness, whereas the baryon and electric charge were treated grand-canonically~\cite{Acharya:2018orn,Sharma:2018jqf}.

The CSM as described above is implemented in the open source  \texttt{Thermal-FIST} package~\cite{Vovchenko:2019pjl}.
All results presented here are obtained using this package.
The annotated macros used to obtain results presented in this paper, as well as the resulting data files, are publicly available~\cite{CSM-github}. 

\section{The roles of different conserved charges in the canonical picture}

The effects of exact charge conservation turn out to be important when the number of particles with the corresponding conserved charge is sufficiently small, typically when it is of the order of unity or less~\cite{Becattini:1995if,Becattini:1997rv,Begun:2004gs}.
The exact conservation of strangeness has commonly been applied only within the so-called strangeness canonical ensemble at both intermediate and at low collision energies as reached e.g. at GSI's SIS and BNL's AGS accelerator facilities~\cite{Cleymans:1990mn,Cleymans:1997sw,Becattini:2000jw,BraunMunzinger:2001as}.
This had been motivated by the small abundances of strange particles produced in those reactions.
The strangeness-canonical picture has recently also been applied for conditions realised at the RHIC-BES~\cite{Adamczyk:2017iwn} and at the LHC~\cite{Acharya:2018orn}.
There, however, the abundance of particles carrying strangeness is not necessarily small as compared to those carrying baryon number and/or electric charge.

\begin{figure}[t]
  \centering
  \includegraphics[width=.47\textwidth]{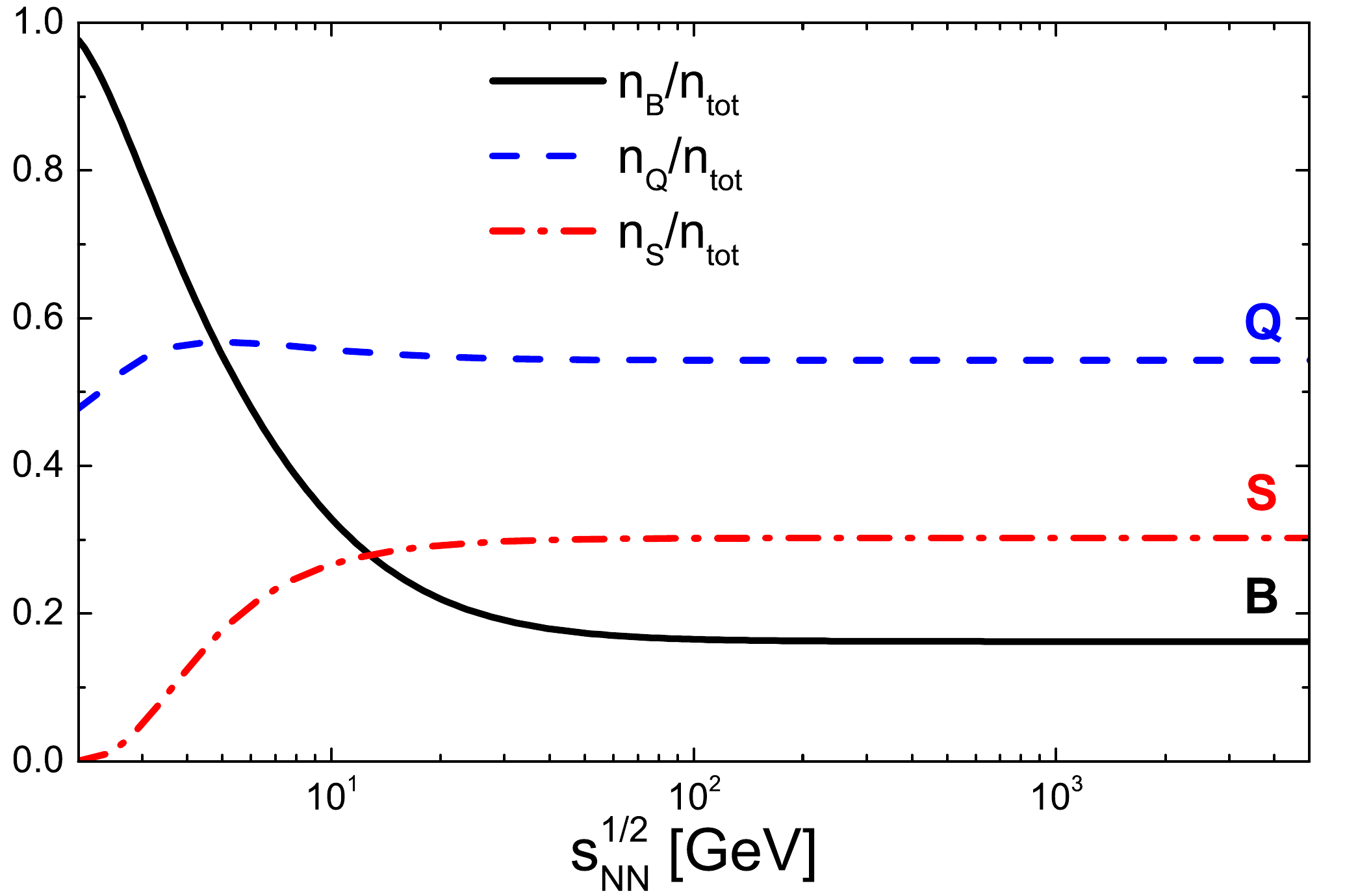}
  \caption{
  Dependence of the hadron number fractions carrying baryon charge~(solid black lines), electric charge~(dashed blue lines), and strangeness~(dash-dotted red lines) on the collision energy, as evaluated in grand-canonical statistical model along the phenomenological chemical freeze-out curve~\cite{Cleymans:2005xv}.
  $n_B$, $n_Q$, and $n_S$ here represent a sum of densities of particles and antiparticles.
  }
  \label{fig:GCEDensities}
\end{figure}

To clarify this question, the relative abundances of particles carrying various conserved charges are evaluated as functions of the collision energy, along the phenomenological chemical freeze-out curve\footnote{Very similar results are obtained when using the updated freeze-out curve from Ref.~\cite{Vovchenko:2015idt}}~\cite{Cleymans:2005xv}.
The results are depicted in Fig.~\ref{fig:GCEDensities}, which confirms the expected small relative abundances of strange hadrons at moderate and at low collision energies, $\sqrt{s_{\rm NN}} \lesssim 10$~GeV.
The strangeness-canonical ensemble is a good starting point for studying the canonical effects at the freeze-out conditions realized at GSI and FAIR, and at the moderate energies of the beam energy scan programs at the SPS and RHIC.
However, the abundance of (anti)baryons is smaller than that of strange hadrons at higher collision energies, $\sqrt{s_{NN}} \gtrsim 10$~GeV.
The role of exact conservation of baryon number at the LHC therefore is expected to be 
even more important than that of strangeness.

\begin{figure*}[t]
  \centering
  \includegraphics[width=.99\textwidth]{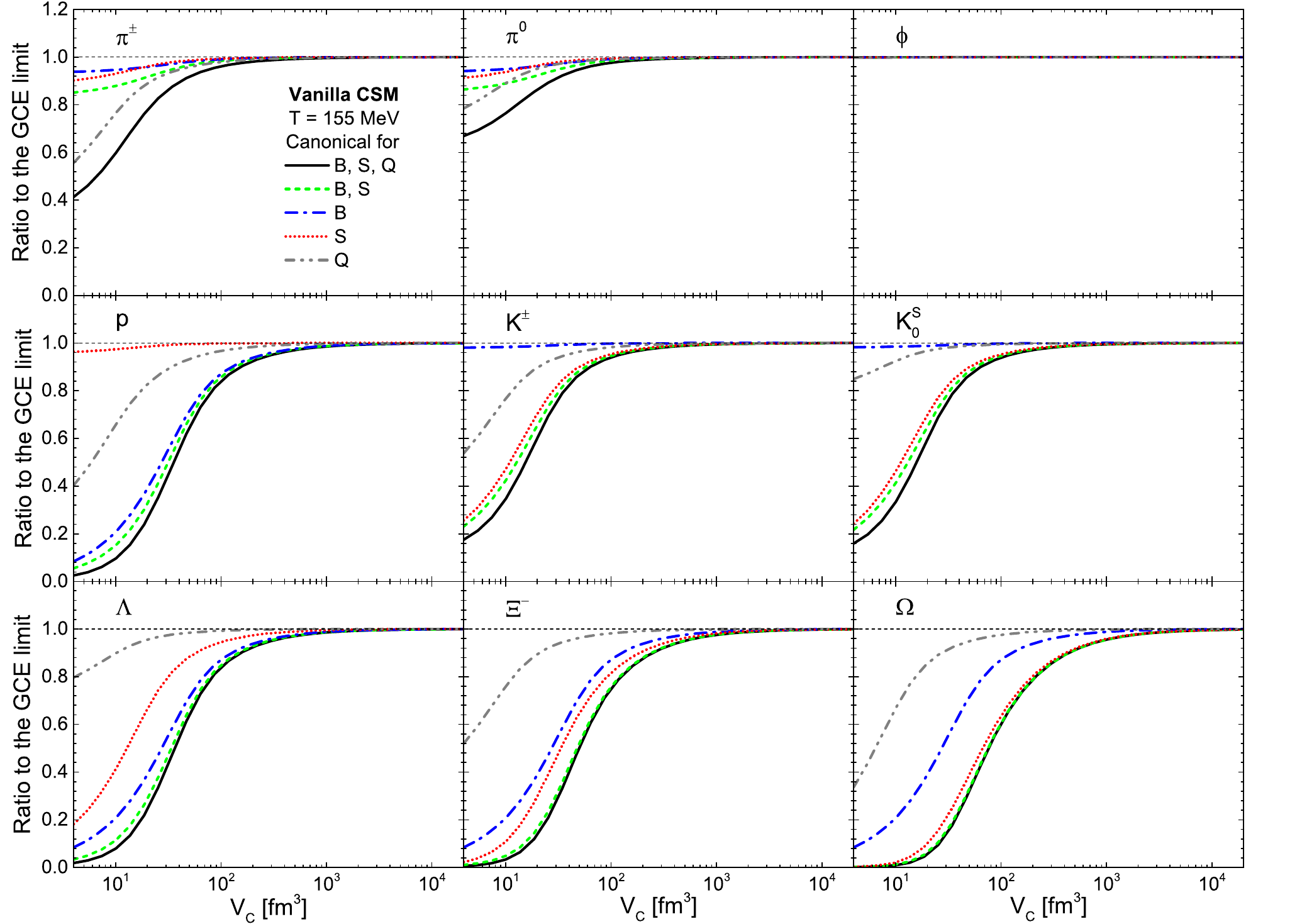}
  \caption{
  Correlation volume dependence of the ratios of various final hadron yields calculated in full canonical~(black lines), baryon-strangeness canonical~(dashed green lines), baryon-canonical~(dash-dotted blue lines), strangeness-canonical~(dotted red lines), and charge-canonical~(dash-double-dotted gray lines) ensembles relative to the limiting grand-canonical values. Calculations are performed at $T = 155$~MeV for zero values of conserved charges, which corresponds to the various systems created in p--p, p--A, and A--A collisions at the LHC.
  }
  \label{fig:CSMYields}
\end{figure*}

The matter created in various colliding systems at the LHC is observed to be baryon-symmetric~\cite{Acharya:2018orn,Abelev:2013haa,Abelev:2013vea},
thus it is characterised by zero values of the conserved net charges.
The role of the various conservation laws at the LHC can best be evaluated from the final yields of various hadron species as obtained in the statistical model with different mixed-canonical ensembles, as a function of the correlation volume $V_c$. 
The computed yields are normalized by the limiting grand-canonical values. 
This is depicted in Fig.~\ref{fig:CSMYields}.
The canonical effects are evidently important for $V_c \lesssim 100-1000$~fm$^3$ at $T = 155$~MeV.
The effects of various conservation laws are different for different hadron species.
The exact conservation of strangeness drives the large canonical suppression effect for the yields of triple-strange $\Omega$ baryons. 
It also has a strong influence on the yields of $\Xi$'s and, to a lesser extent, kaons.
The strangeness conservation is subdominant for other yields.
Exact baryon number conservation is most important for the yields of protons and $\Lambda$'s. It also yields a sizable influence on the yields of $\Xi$'s.
The exact conservation of the electric charge is important for pions.
Even the final yields of $\pi^0$ are affected by the canonical effects.
Even though $\pi^0$ is a neutral particle, the yields of $\pi^0$ receive sizable feeddown contribution from unstable resonances, which are affected by the canonical suppression.
The final yields of charged pions are affected by the baryon and strangeness conservation for the same reason.
The yields of $\phi$ mesons are found to be unaffected by any canonical effects, as the $\phi$ meson is a neutral particle with no known feeddown contributions from non-neutral particles.

These results prove that only the simultaneous canonical treatment of all three conserved charges is sufficient for a quantitative CSM application to the LHC data for $V_c \lesssim 100$~fm$^3$. 
A grand-canonical treatment of the electric charge might be permitted for values of $V_c \gtrsim 100$~fm$^3$.
We do not find such conditions at the LHC where a strangeness-only canonical treatment of the relevant hadron yields is appropriate.

In the following, we apply the CSM with the simultaneous exact conservation of all three conserved charges: the baryon number~$B$, the electric charge~$Q$, and strangeness~$S$.

\section{Vanilla CSM at the LHC}
\label{sec:VanillaCSM}

\begin{figure*}[!ht]
  \centering
  \includegraphics[width=.99\textwidth]{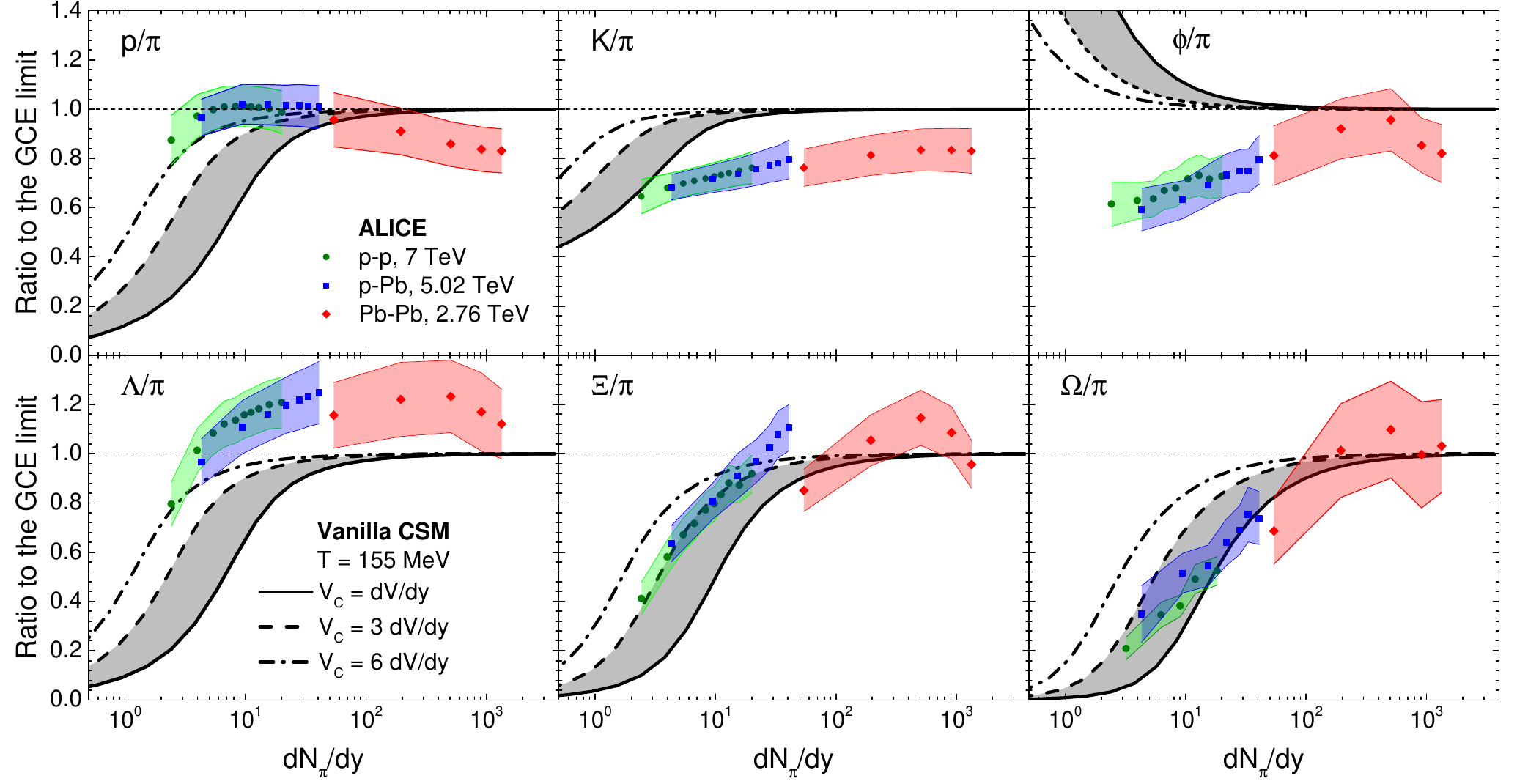}
  \caption{
  The ratios of various final hadron-to-pion yields are plotted versus charged the pion multiplicity 
  as evaluated in the vanilla CSM with exact conservation of baryon number, electric charge, and strangeness.
  The green circles, blue squares, and red diamonds depict the corresponding ratios as measured by the ALICE collaboration at the LHC in p-p~(7~TeV), p-Pb~(5.02~TeV), and Pb-Pb~(2.76~TeV) collisions, respectively.
  Both the calculated results and the data are scaled by the grand-canonical limiting values as evaluated in the CSM at $T = 155$~MeV for $\mu_B = 0$.
  }
  \label{fig:CSMYieldRatios}
\end{figure*}

\subsection{Correlation volume}
\label{sec:flucs}

The matter produced and observed at the LHC in the central rapidity region is nearly baryon-symmetric, even on an event-by-event basis.
Hence, we do employ the canonical ensemble method using zero net values of all conserved charges, i.e. $B = Q = S = 0$.

To apply the CSM to experimental data one needs to relate the mean multiplicities $\langle N^{\rm ce}_j \rangle^{\rm tot}$ calculated in the CSM to the rapidity densities, $dN_j / dy$, measured experimentally.
The connection between the correlation volume $V_c$ and the volume $dV / dy$ which corresponds to one unit of rapidity needs to be established.
This entails a conceptual issue for the approach: all midrapidity slices of the system are open systems, where net values of conserved charges fluctuate from one event to another. 
Hence, there is no reason to enforce $V_c = dV / dy$.
On the other hand, as particle production processes are typically localized in space microscopically, a consideration of local conservation of charges, which translates to localized regions in the longitudinal rapidity~\cite{Castorina:2013mba}, seems feasible.
Our earlier work~\cite{Vovchenko:2018fiy} has varied $V_c$ between one and few units of rapidity, i.e. $V_c = k \,dV / dy$ with $k \geq 1$. 
This choice is suggested by consideration of a causal connection of fireballs which populate the longitudinal rapidity space~\cite{Castorina:2013mba}.

It is argued here that estimates for the value of $k$ can be obtained from measurements of net-proton fluctuations in high-energy reactions at the LHC.
Fluctuation measurements are affected by conservation laws~\cite{Jeon:2000wg,Begun:2004gs,Bzdak:2012an,Braun-Munzinger:2016yjz,Pruneau:2019baa}.
These lead to deviations from the independent particle production baseline~(Poisson statistics).
For instance, assuming that net-proton fluctuations measured in a particular acceptance are affected by the exact conservation of baryon number in a certain correlation volume, whereas the role of other conservation laws and of dynamical effects is negligible, one obtains the following ratio $\kappa_2(\text{p}-\bar{\text{p}}) / ( \langle \text{p} \rangle + \langle \bar{\text{p}} \rangle )$ of variance of net-proton fluctuations to the mean number of protons and antiprotons at the LHC~\cite{Jeon:2000wg,Braun-Munzinger:2016yjz}:
\eq{\label{eq:kappacons}
\frac{\kappa_2(\text{p}-\bar{\text{p}})}{\langle \text{p} \rangle + \langle \bar{\text{p}} \rangle } = 1 - \alpha.
}
Here $\alpha$ is the ratio of the mean number of protons $\langle p \rangle$ in the acceptance window where the net-proton fluctuations are measured relative to the mean number of $B = + 1$ baryons in the correlation volume across which the exact baryon number conservation is enforced.

The quantity $\kappa_2(\text{p}-\bar{\text{p}}) / ( \langle \text{p} \rangle + \langle \bar{\text{p}} \rangle )$ has been measured by the ALICE collaboration in Pb--Pb collisions at $\sqrt{s_{\rm NN}} = 2.76$~TeV for various centralities.
Preliminary data have been presented in Ref.~\cite{Rustamov:2017lio} for the acceptance window $0.6 < p < 1.5$~GeV/$c$ in momentum and $|\eta| < 0.8$ in pseudorapidity.
The measured ratio values are below unity, which has been attributed to the baryon number conservation~\cite{Rustamov:2017lio}.
To connect these measurements with the assumed local baryon conservation\footnote{The consideration of \emph{local} baryon number conservation is one difference to prior studies~\cite{Braun-Munzinger:2016yjz,Rustamov:2017lio} where a \emph{global} conservation was studied instead.} in a correlation volume $V_c = k \, dV/dy$ around midrapidity we assume that longitudinal boost invariance holds in the rapidity region $-k/2 < y < k/2$.
This is a reasonable approximation as long as one stays within a few units around midrapidity. 
These considerations can be further improved by using measurements of balance functions~\cite{Bass:2000az,Pratt:2011bc}, as discussed recently in Ref.~\cite{Pruneau:2019baa}.
The parameter $\alpha$ is given by
\eq{\label{eq:alphalocalcons}
\alpha = \frac{\langle \text{p} \rangle}{k \, d N_B / dy},
}
Here $d N_B / dy$ is the rapidity density of the $B = +1$ baryons, thus $k \, d N_B / dy$ is the mean number of baryons in the correlation volume $V_c$.
Our statistical model estimate suggests that the final state protons correspond to approximately a third of all final state baryons at the LHC, the rest being equally distributed between neutrons and hyperons.
Thus, $d N_B / dy \approx 3 \, d N_{\text{p}} / dy$.
The value of $k$ can therefore be estimated from 
Eq.~\eqref{eq:alphalocalcons}
for a given centrality from the measured values of $\alpha \equiv 1 - \kappa_2(\text{p}-\bar{\text{p}}) / ( \langle \text{p} \rangle + \langle \bar{\text{p}} \rangle )$, $\langle \text{p} \rangle$, and $d N_{\text{p}} / dy$:
\eq{\label{eq:k}
k \approx \frac{\langle \text{p} \rangle}{3\,\alpha\,d N_{\text{p}} / dy}.
}
For the most peripheral available Pb--Pb bin~(60-70\%) one has
$\alpha \simeq 0.07$
and $\langle \text{p} \rangle \simeq 1.14$~\cite{Rustamov:2017lio}, and $d N_{\text{p}} / dy \simeq 1.9$~\cite{Abelev:2013vea}, which gives $k \simeq 3$.
Analysis of other centrality bins gives values in the range $k \sim 3-4$, with the exception of the two most central bins~(0-5\% and 5-10\%) where higher values of $k \sim 5-6$ are indicated by the data, suggesting that local conservation assumption likely approaches the global conservation assumption for larger systems.
The analysis suggests that $k \sim 3-6$ might be a reasonable estimate for the canonical correlation volume.
A detailed analysis of the upcoming fluctuation data from the ALICE collaboration will be reported elsewhere, once the net-proton fluctuation data are finalized.

\subsection{Assumptions}

The simplest version of the CSM applied to the LHC data, henceforth dubbed the "Vanilla CSM", is based on the following scenario:
\begin{itemize}
    \item the full chemical equilibrium is established at the chemical freeze-out stage.
    \item a constant chemical freeze-out temperature of $T = 155$~MeV exists across all multiplicity bins, as suggested by the statistical model fits to the hadron yield data in most central Pb--Pb collisions.
    \item the multiplicity dependence of various hadron yield ratios is driven by the canonical suppression only, i.e. by the changing value of $V_c$.
    \item the correlation volume in rapidity is varied between $V_c = dV / dy$ and $V_c = 6\, dV / dy$.
\end{itemize}

\subsection{Results}

Figure~\ref{fig:CSMYieldRatios} depicts the ratios $p/\pi$, $K/\pi$, $\phi / \pi$, $\Lambda / \pi$, $\Xi / \pi$, and $\Omega / \pi$, evaluated as a function of $d N_\pi / dy$, using the vanilla CSM for $V_c = dV / dy$, $V_c = 3 \, dV/dy$, and $V_c = 6 \, dV/dy$.
Both the calculated values and the data are normalized by the limiting grand-canonical values of the ratios as calculated in the CSM.
Figure~\ref{fig:CSMYieldRatios} does reflect the limited level of agreement of the statistical model with the data. This is true also in the limit of high multiplicities, i.e. for central Pb--Pb collisions.
This is different from the strangeness-canonical ensemble analysis presented in Ref.~\cite{Acharya:2018orn}, where the data and the model results were normalized separately by the high-multiplicity limiting values in the data and in the model, respectively.

The vanilla CSM provides a fair description of the $\Lambda / \pi$, $\Xi / \pi$, and $\Omega / \pi$ ratios. The trend of a decreasing $K/\pi$ ratios at small multiplicities is also captured, although the model overshoots the data rather severely.
The description of the $K/\pi$ ratio worsens as multiplicity is decreased.
The $p/\pi$ ratio is strongly affected by the canonical suppression in the model. 
The data, on the other hand, shows a suppression of $\text{p}/\pi$ at high multiplicity, but no conclusive evidence for a suppression of $\text{p}/\pi$ in small systems, except for the lowest multiplicity bins in p--p collisions.
The vanilla CSM agrees with the data only if a rather large correlation volume $V_c \simeq 6 \, dV / dy$ value is used, 
which is not supported by the observed yields of strange hadrons.

The vanilla CSM had been used to study the multiplicity dependence of the yields of light nuclei at the LHC~\cite{Vovchenko:2018fiy}.
The CSM predicts the suppression of the ratios $\text{d}/\text{p}$, $^3\text{He}/\text{p}$, $^3_\Lambda \text{H} / \text{p}$, and $^4\text{He}/\text{p}$ at the lower multiplicities.
A fair description of the available data on $d/\text{p}$ and $^3\text{He}/\text{p}$ was achieved for $V_c = 3 \, dV/dy$.
This result puts the vanilla CSM in a certain tension with the $\text{p}/\pi$ ratio data: $p/\pi$ is notably suppressed in the model for $V_c = 3 \, dV/dy$ already for multiplicities which do not show a suppression in the data.

A severe problem is seen in Fig.~\ref{fig:CSMYieldRatios} when describing the data of the $\phi / \pi$ ratio: 
the $\phi$ meson yield is unaffected by canonical suppression, as $\phi$ is a neutral meson. 
The yields of pions, however, are suppressed by the canonical suppression~(see Fig.~\ref{fig:CSMYields}), which leads to a predicted strong increase of the $\phi / \pi$ ratio towards smaller multiplicities.
However, the data show just the opposite: the $\phi / \pi$ ratios are smaller at smaller multiplicities.
Unless the production mechanism for $\phi$ is completely different from all other hadrons, this invalidates the vanilla CSM picture in p--p and p--Pb collisions.

\section{CSM with incomplete equilibration of strangeness and multiplicity dependent temperature}

\subsection{Considerations}

The relatively simple vanilla CSM successfully describes certain features of the measured yield data at the LHC. 
More involved considerations are necessary in light of the tensions of the model with the data on $p/\pi$, $K/ \pi$ and, in particular, the $\phi/\pi$ ratio~(see Fig.~\ref{fig:CSMYieldRatios}).

Consider a possibility that the chemical freeze-out temperature is not constant across all multiplicities, but that smaller systems might be characterised by larger chemical freeze-out temperatures.
This would correspond to an earlier decoupling of the inelastic hadronic reactions, e.g. stemming from larger radial flow gradients in smaller collision systems~\cite{Shuryak:2013ke}.
This type of a scenario is observed for the decoupling of the (pseudo)elastic hadronic reactions, characterised by the kinetic freeze-out temperature values extracted from the blast-wave fits to the $p_T$ spectra of pions, kaons, and protons~\cite{Abelev:2013vea,Acharya:2018orn}. 
The larger $T_{\rm kin}$ values observed in p--p collisions~($T_{\rm kin} \sim 170$~MeV) as compared to central Pb--Pb collisions~($T_{\rm kin} \sim 100$~MeV) seem to indicate an earlier decoupling of these reactions in small systems, which usually also implies a shorter lifetime of the hadronic phase. 

The suppression of the $\phi/\pi$  ratio observed in smaller collision systems, as well as the severe overestimation of the data on the $K/\pi$ ratio, may indicate incomplete 
chemical equilibrium in the thermal-statistical picture.
One possible remedy is a multiple chemical freeze-out scenario with an earlier freeze-out of strangeness~\cite{Bellwied:2013cta,Chatterjee:2016cog}. 
The present work incorporates the incomplete equilibration of strangeness into the CSM by introducing the strangeness saturation factor $\gamma_S$~\cite{Koch:1986ud,Rafelski:1991rh}, which results in the following modification of the one-particle partition function in Eq.~\eqref{eq:zj}: $z_j^n \to \gamma_S^{|s_j|} \, z_j^n$, $|s_j|$ being the number of strange quarks and antiquarks in the quark content of particle species $j$. 
The $\phi$ meson is a particularly interesting species in this regard: $\phi$ is a neutral particle with zero net strangeness which is unaffected by the exact conservation of strangeness in the canonical suppression picture.
In the strangeness non-equilibrium picture, however, $\phi$ is a double-strange particle consisting of a strange quark-antiquark pair.
This is similar to fragmentation models~\cite{Sjostrand:2007gs,Skands:2014pea,Bierlich:2014xba,Pierog:2013ria}, 
where coalescence of two strangeness carrying strings is needed to form $\phi$, making this meson behave as effectively a double-strange particle.
The experimentally observed suppression of the $\phi/\pi$ ratio suppression at small multiplicities suggests an introduction of a multiplicity-dependent $\gamma_S \leq 1$.

\begin{figure*}[!ht]
  \centering
  \includegraphics[width=.80\textwidth]{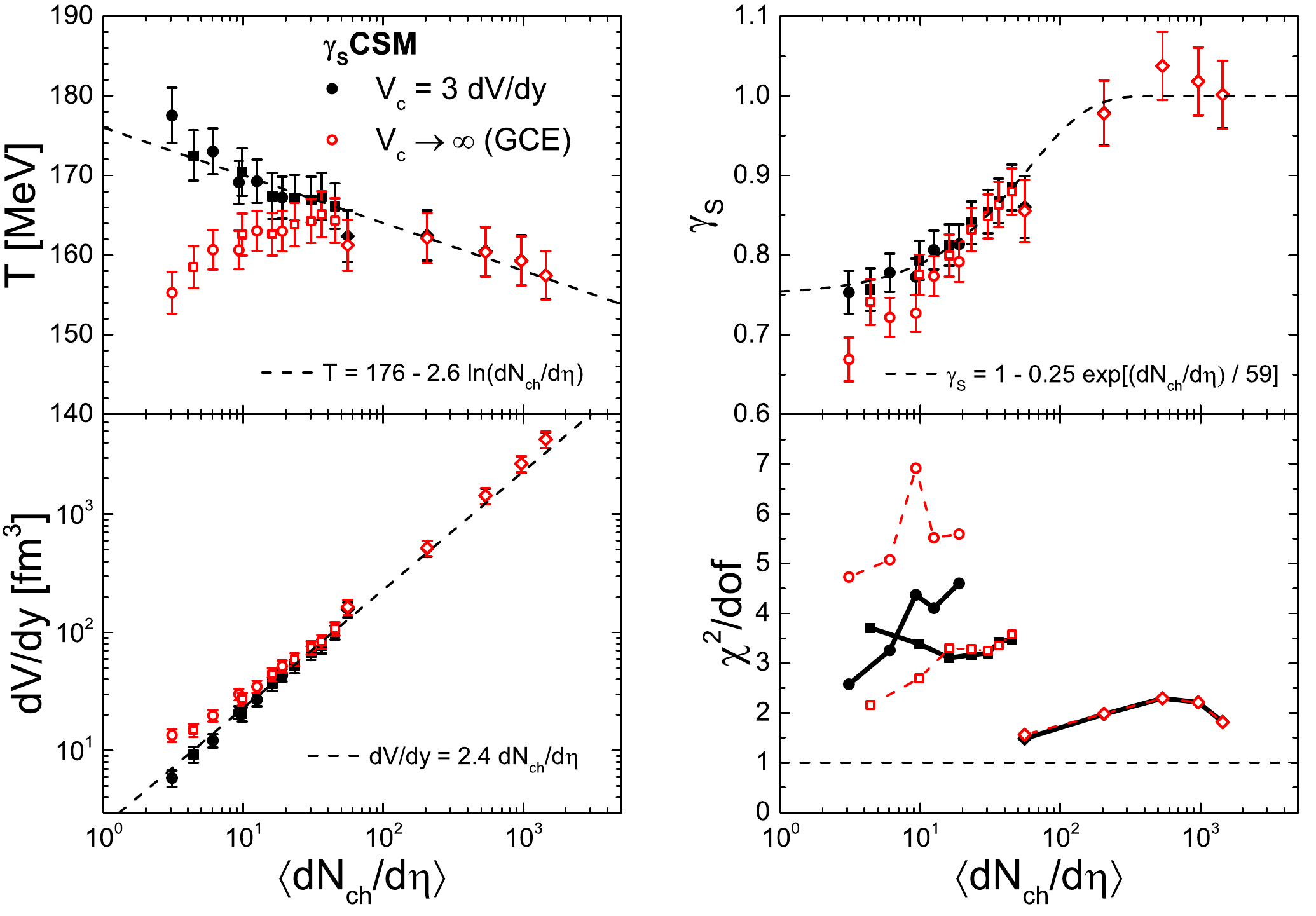}
  \caption{
  Results of the thermal fits to p-p~(circles), p-Pb~(squares), and Pb-Pb~(diamonds) data of the ALICE collaboration within the CSM with $V_c = 3 \, dV/dy$~(full black symbols) and in the grand-canonical limit $V_c \to \infty$~(open red symbols).
  The dependence of the chemical freeze-out temperature $T$~(a), the strangeness saturation parameter $\gamma_S$~(b), the volume per unit rapidity $dV/dy$~(c), and $\chi^2/dof$~(d) on the charged particle multiplicity $\langle d N_{\rm ch}/d \eta \rangle_{|\eta|<0.5}$ is depicted.
  The lines in (d) connect the extracted $\chi^2/dof$ values in different multiplicity bins in different systems to guide the eye.
  }
  \label{fig:CSMFitParams}
\end{figure*}

\subsection{The data set and fitting procedure}

The model which includes the two last considerations above, termed $\gamma_S$CSM,
is used to perform thermal fits to the yields of $\pi$, $K$, $K_0^S$, $\phi$, p, $\Lambda$, $\Xi$, and $\Omega$, as measured by the ALICE collaboration in different centrality bins at 7~TeV p--p~\cite{Acharya:2018orn}, 5.02~TeV p--Pb~\cite{Abelev:2013haa,Adam:2015vsf,Adam:2016bpr}, and 2.76~TeV Pb--Pb~\cite{Abelev:2013vea,Abelev:2013xaa,ABELEV:2013zaa,Abelev:2014uua} collisions.
The yields of particles and antiparticles are symmetrized,
i.e. mean values of particle and antiparticle yields are taken.
The following centrality bins are considered: V0M I-II, III-IV, V-VI, VII-VIII, IX-X in p--p, V0A 0-5\%, 5-10\%, 10-20\%, 20-40\%, 40-60\%, 60-80\%, and 80-100\% in p--Pb, and 0-10\%, 10-20\%, 20-40\%, 40-60\%, and 60-80\% in Pb--Pb.
The centrality binning for the $\phi$'s in p--p is different from the binning of other yields~(see Table~12 in Ref.~\cite{Acharya:2018orn}).
The $\phi$ yields are reconstructed in the needed centrality bins from the neighbouring centrality bins through a linear interpolation in charged multiplicity.
The parameters fitted are: the temperature, $T$, the volume per rapidity unit, $dV/dy$, and the strangeness saturation parameter, $\gamma_S$.
All conserved charges in the CSM fit are fixed to zero: $B = Q = S = 0$.
Two cases are considered: $V_c = 3 \, dV/dy$ and $V_c \to \infty$. 
The latter case corresponds to the grand-canonical statistical model calculation while in the former case a local exact charge conservation in a volume suggested by measurements of net-proton fluctuations~(see Sec.~\ref{sec:flucs}) is enforced.

A similar analysis, including $\gamma_S$, has recently been presented in Ref.~\cite{Sharma:2018jqf}.
There are two important differences there compared to the present work: 
(i) in Ref.~\cite{Sharma:2018jqf} the $\phi$ meson yields were excluded from the fits, here they are included; 
(ii) in Ref.~\cite{Sharma:2018jqf} the canonical correlation volume was forced to be equal to the volume per one unit of rapidity, $V_c = dV/dy$, here this is not the case. In fact, it is the $V_c = dV/dy$ constraint which appears to be the primary reason why the analysis of Ref.~\cite{Sharma:2018jqf} has difficulties to accommodate the $\phi$ meson yields in the $\gamma_S$CSM picture.

\begin{figure*}[t]
  \centering
  \includegraphics[width=.32\textwidth]{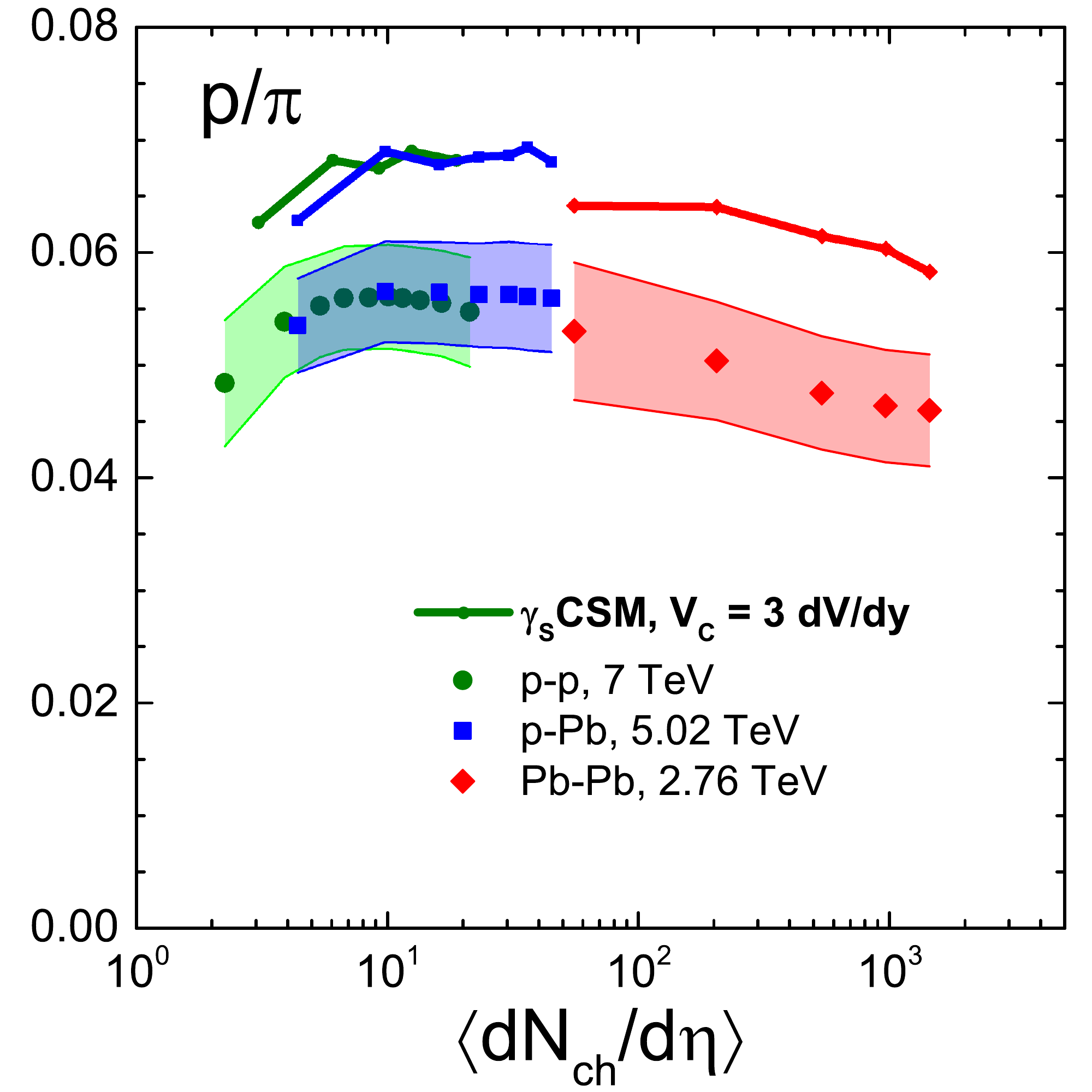}
  \includegraphics[width=.32\textwidth]{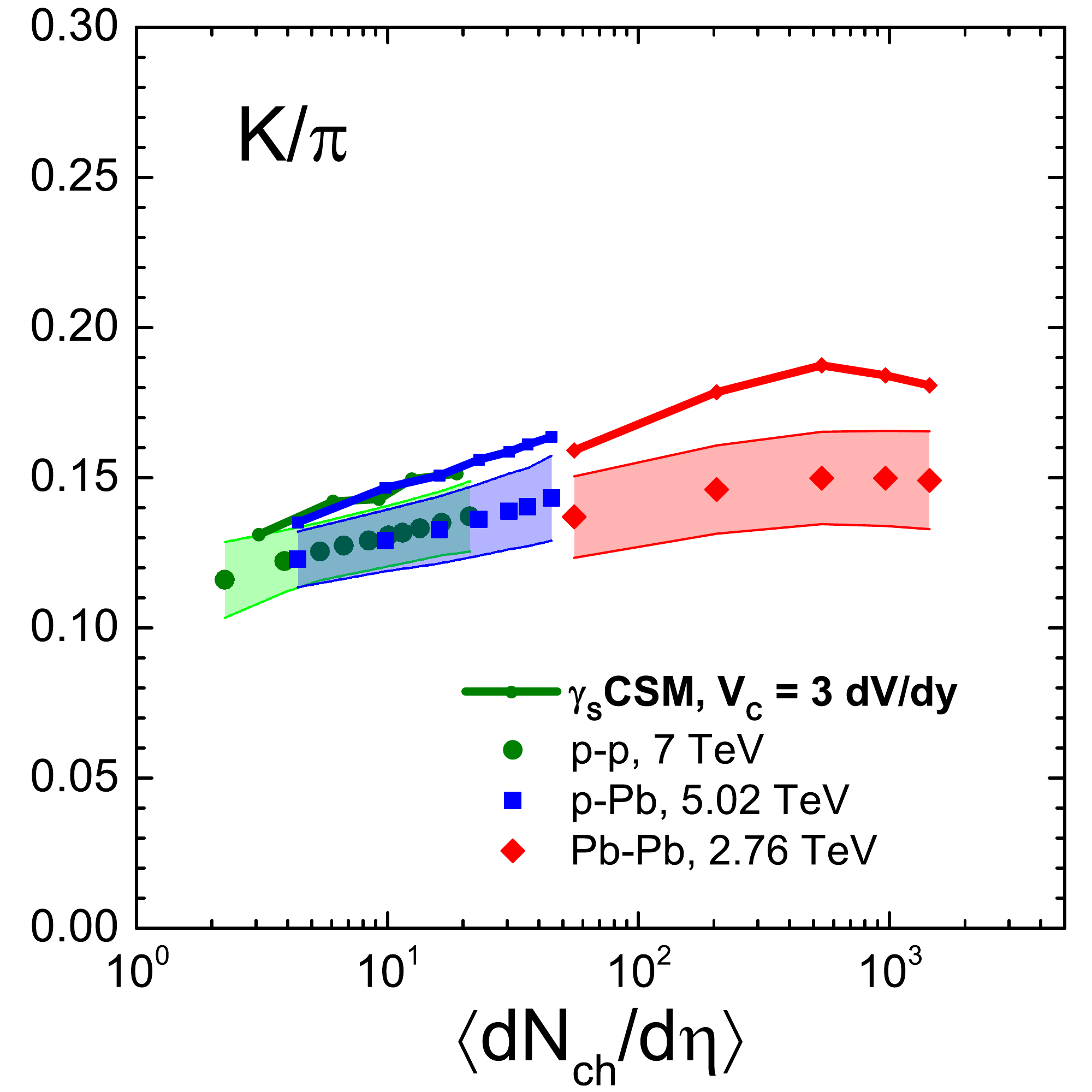}
  \includegraphics[width=.32\textwidth]{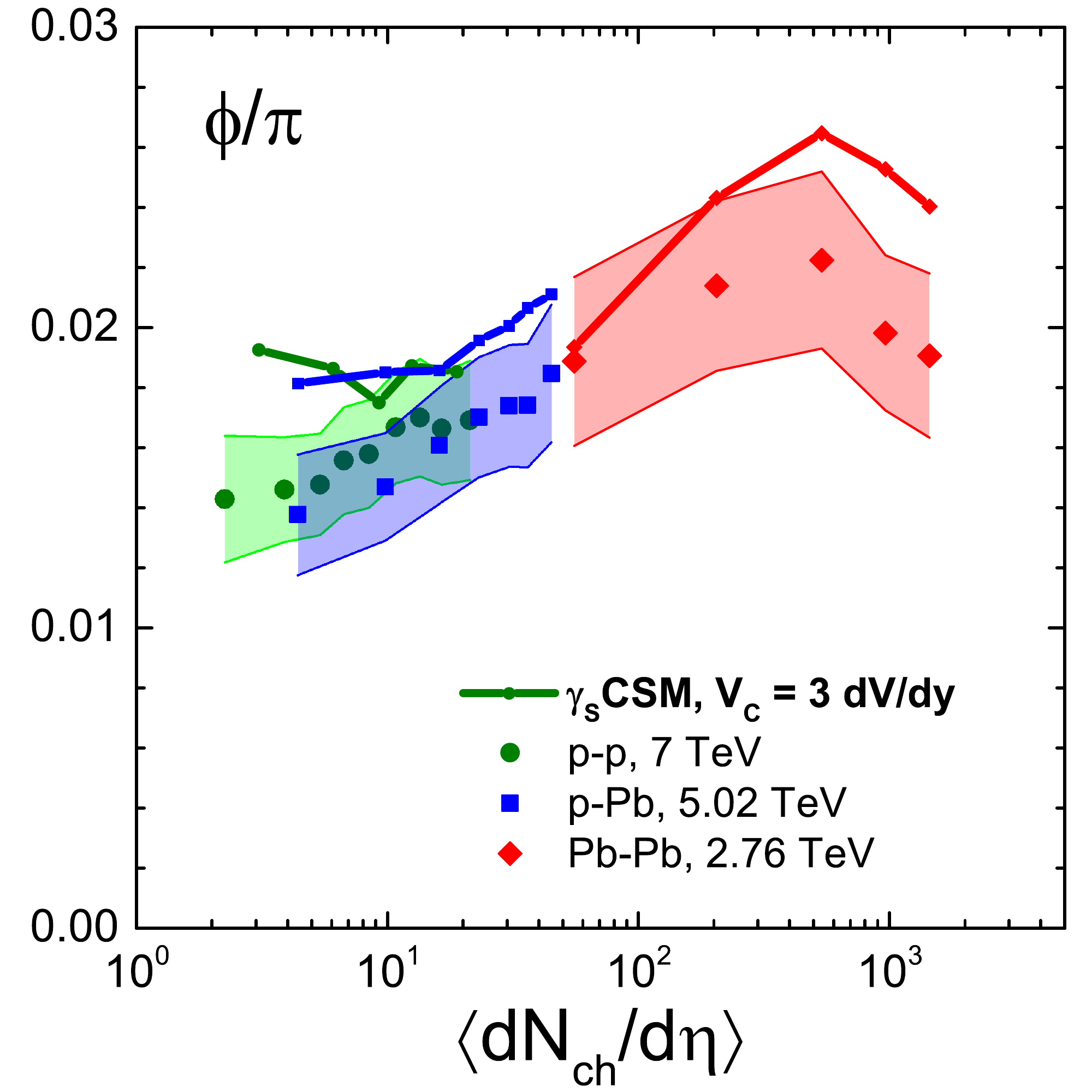}
  \includegraphics[width=.32\textwidth]{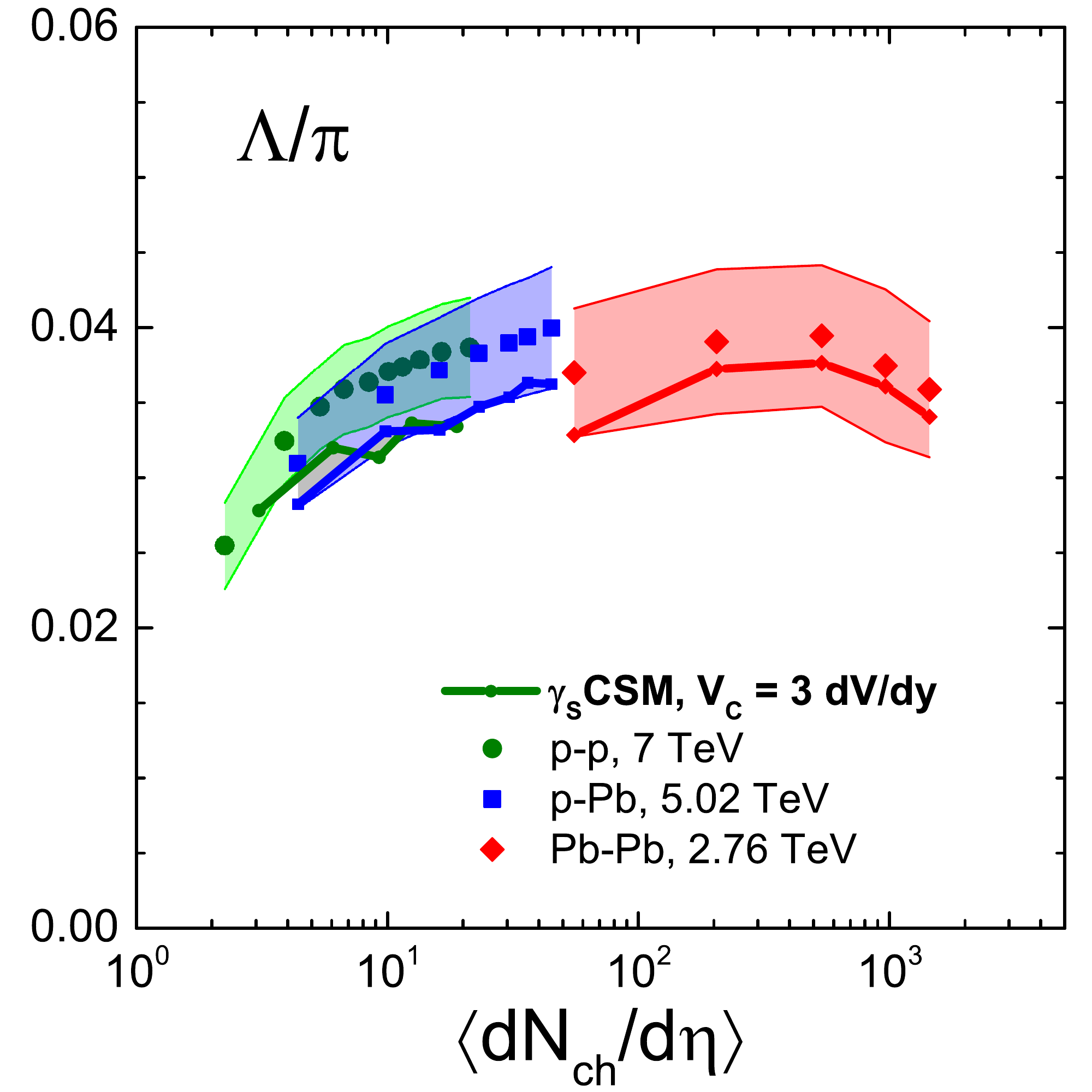}
  \includegraphics[width=.32\textwidth]{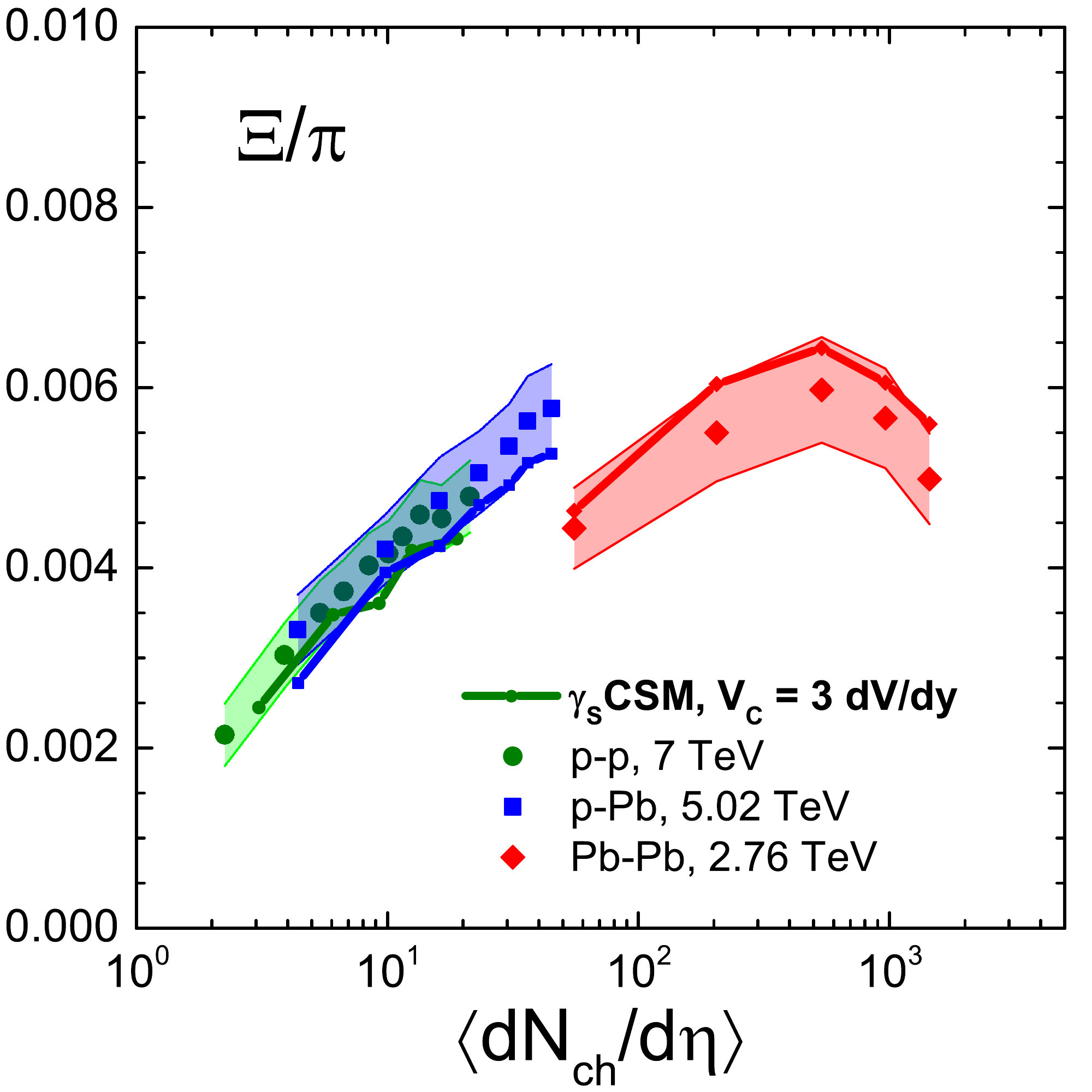}
  \includegraphics[width=.32\textwidth]{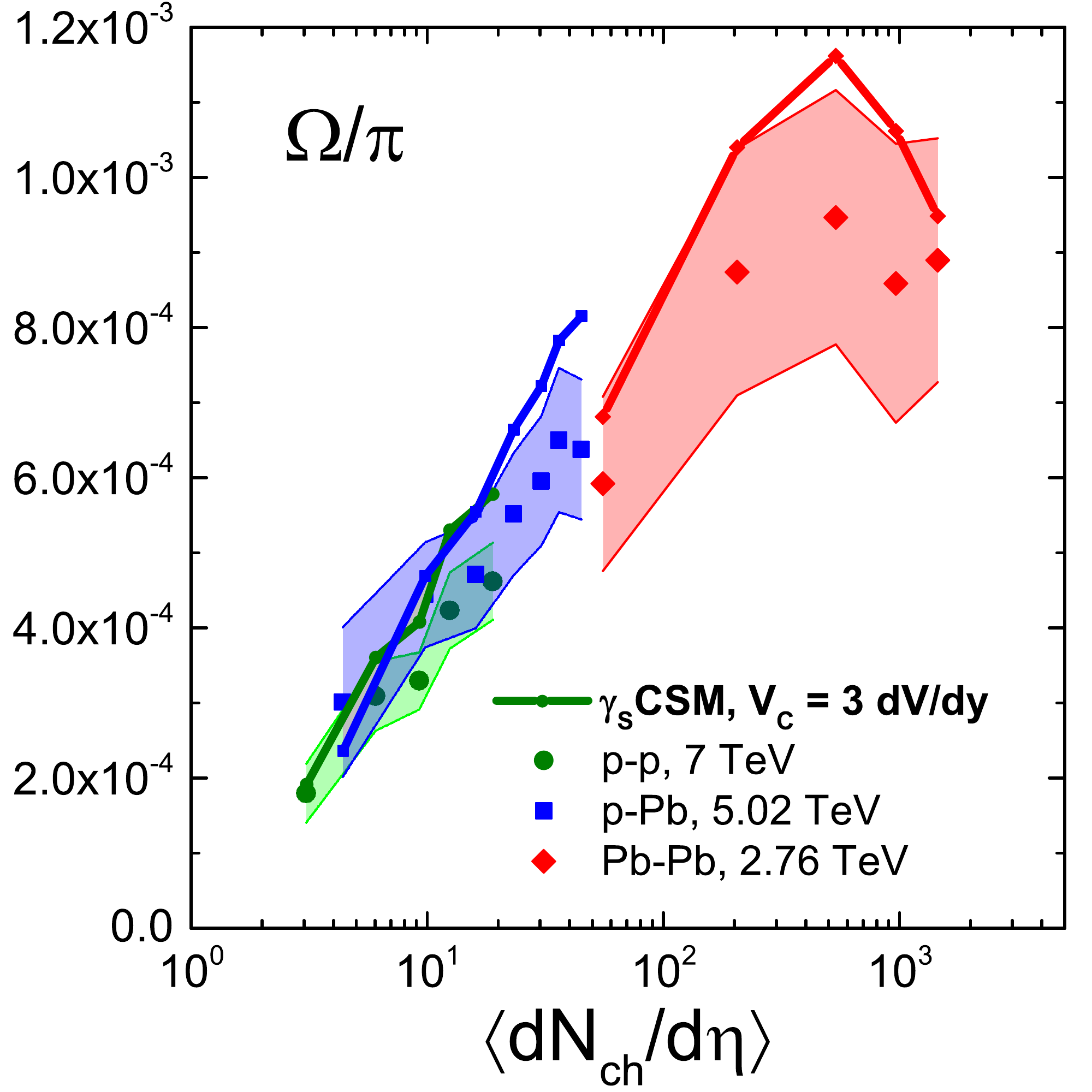}
  \caption{
  The dependence of yield ratios (a) $p/\pi$, (b) $K/\pi$, (c) $\phi / \pi$, (d) $\Lambda / \pi$, (e) $\Xi/\pi$, and (f) $\Omega / \pi$ on the charged particle multiplicity $\langle d N_{\rm ch} / d \eta \rangle_{|\eta| < 0.5}$, evaluated in the $\gamma_S$CSM with $V_c = 3 \, dV/dy$ for the thermal parameters extracted for each multiplicity bin through the $\chi^2$ minimization.
  The green circles, blue squares, and red diamonds depict the corresponding ratios measured by the ALICE collaboration in p-p~(7~TeV), p-Pb~(5.02~TeV), and Pb-Pb~(2.76~TeV) collisions, respectively, the bands depicting the corresponding experimental uncertainties.
  }
  \label{fig:CSMFittedYields}
\end{figure*}

\subsection{The extracted parameters}

The values of the extracted parameters are depicted in Fig.~\ref{fig:CSMFitParams} as a function of the charged multiplicity $\langle d N_{\rm ch} / d \eta \rangle_{|\eta| < 0.5}$.
For $\langle d N_{\rm ch} / d \eta \rangle_{|\eta| < 0.5} \gtrsim 50$ the results between the $\gamma_S$CSM and the grand-canonical $\gamma_S$SM are almost identical, indicating that the canonical effects are irrelevant for very large systems.
There are important differences at $\langle d N_{\rm ch} / d \eta \rangle_{|\eta| < 0.5} \lesssim 50$:
\begin{itemize}
    \item The fit quality in p--p  is systematically better for p--p within the canonical ensemble than within the GCE. 
    On the other hand, fits to the p--Pb data do not indicate a preference of the canonical ensemble relative to the GCE. 
    In fact, the data description in the lowest two multiplicity bins in p--Pb is superior for GCE than for the CE. 
    The apparent differences between p--p and p--Pb appear to be related to the data quality of the measured $\Omega$ yields: in p--p collisions the experimental uncertainties are notably smaller than in p--Pb. 
    If $\Omega$ yields are removed from the p--p fits, then no preference of the canonical picture over the grand-canonical one is seen. 
    We conclude that $\Omega$ yields are a sensitive probe to distinguish the effects of canonical suppression from an incomplete equilibration of strangeness. 
    New and accurate measurements of $\Omega$ yields in p--p and p--Pb will thus be important in that regard.

    \item For $V_c = 3 \, dV/dy$ the extracted temperature decreases monotonically with $d N_{\rm ch} / d \eta$, from the maximum value of $T \simeq 175$~MeV for the lowest multiplicity bins in p--p collisions to the minimum value of $T \simeq 155$~MeV in the most central Pb--Pb collisions.
    This is in line with a possible earlier chemical freeze-out in smaller systems as discussed above.
    In the GCE picture, the extracted temperature behaves in the opposite way, $T$ is the smallest for the smallest multiplicity bins.
    
    \item The strangeness saturation parameter $\gamma_S$ is a monotonically increasing function of $d N_{\rm ch} / d \eta$, reaching $\gamma_S \simeq 1$ at $d N_{\rm ch} / d \eta \gtrsim 100-200$.
    For p--p and p--Pb the values of $\gamma_S$ in the GCE analysis are somewhat smaller than those in the CE analysis, in particular for $d N_{\rm ch} / d \eta \lesssim 10$. These smaller values of $\gamma_S$ may mimick the canonical suppression of strangeness in the GCE picture.
    
    \item $dV/dy$ is a monotonically increasing function of $d N_{\rm ch} / d \eta$. 
    The CE fits are described fairly well with a linear dependence, $dV/dy~[\text{fm}^3] \simeq 2.4 \, d N_{\rm ch} / d \eta$, as shown by the dashed black line in Fig.~\ref{fig:CSMFitParams}.
    The $dV/dy$ dependence on $d N_{\rm ch} / d \eta$ extracted from the GCE fits, on the other hand, does not show a linear dependence if the whole range of $d N_{\rm ch} / d \eta$ values is considered.
    
\end{itemize}

Overall, the behaviour of the extracted parameters gives a fairly consistent picture in the $\gamma_S$CSM-case with $V_c = 3 \, dV/dy$.
We provide the charged multiplicity dependence of the extracted parameters in a parametrized form.
The parametrization corresponds to a multiplicity range $3 \lesssim d N_{\rm ch} / d \eta \lesssim 1500$.

The chemical freeze-out temperature is given by
\eq{\label{eq:Tparam}
T = T_0 - \Delta T \, \ln(d N_{\rm ch} / d \eta),
}
with $T_0 = (176 \pm 1)$~MeV  and $\Delta T = (2.6 \pm 0.2)$~MeV.

The strangeness saturation parameter $\gamma_S$ is
\eq{\label{eq:gsparam}
\gamma_S = 1 - A \, \exp \left[- \frac{d N_{\rm ch} / d \eta}{B} \right].
}
Here $A = 0.25 \pm 0.01$ and $B = 59 \pm 6$.

Finally, as already pointed out above, the volume parameter $dV/dy$ is linearly proportional to $d N_{\rm ch} / d \eta$:
\eq{\label{eq:dVdyparam}
dV/dy = v \, d N_{\rm ch} / d \eta, \qquad v = (2.4 \pm 0.2)~\text{fm}^3.
}

We have additionally considered $\gamma_S$CSM fits with $V_c = dV/dy$. 
The extracted temperature exhibits even larger values at small multiplicities, reaching $T_{\rm ch} \sim 200$~MeV for the lowest p--p bin, in agreement with the results reported in~Ref.~\cite{Sharma:2018jqf}. 
We find that the canonical effects in the $V_c = dV/dy$ case are too strong in p--p and p--Pb collisions, leading to a significant worsening~($\chi^2/{\rm dof} \sim 5-20$) of the data description quality at small multiplicities $dN_{\rm ch} / d \eta \lesssim 10$.

\subsection{Hadron yield ratios}

\begin{figure*}[t]
  \centering
  \includegraphics[width=.32\textwidth]{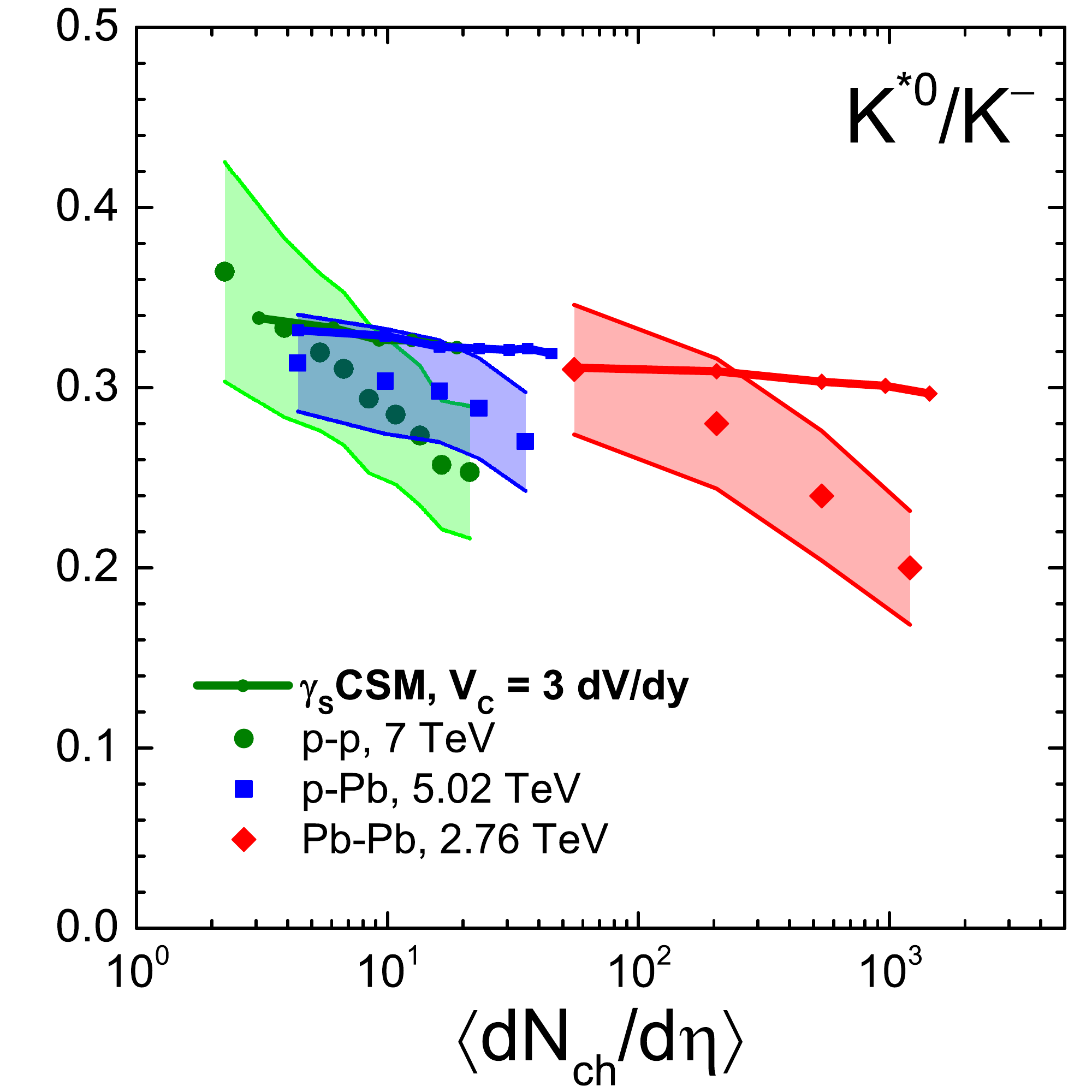}
  \includegraphics[width=.32\textwidth]{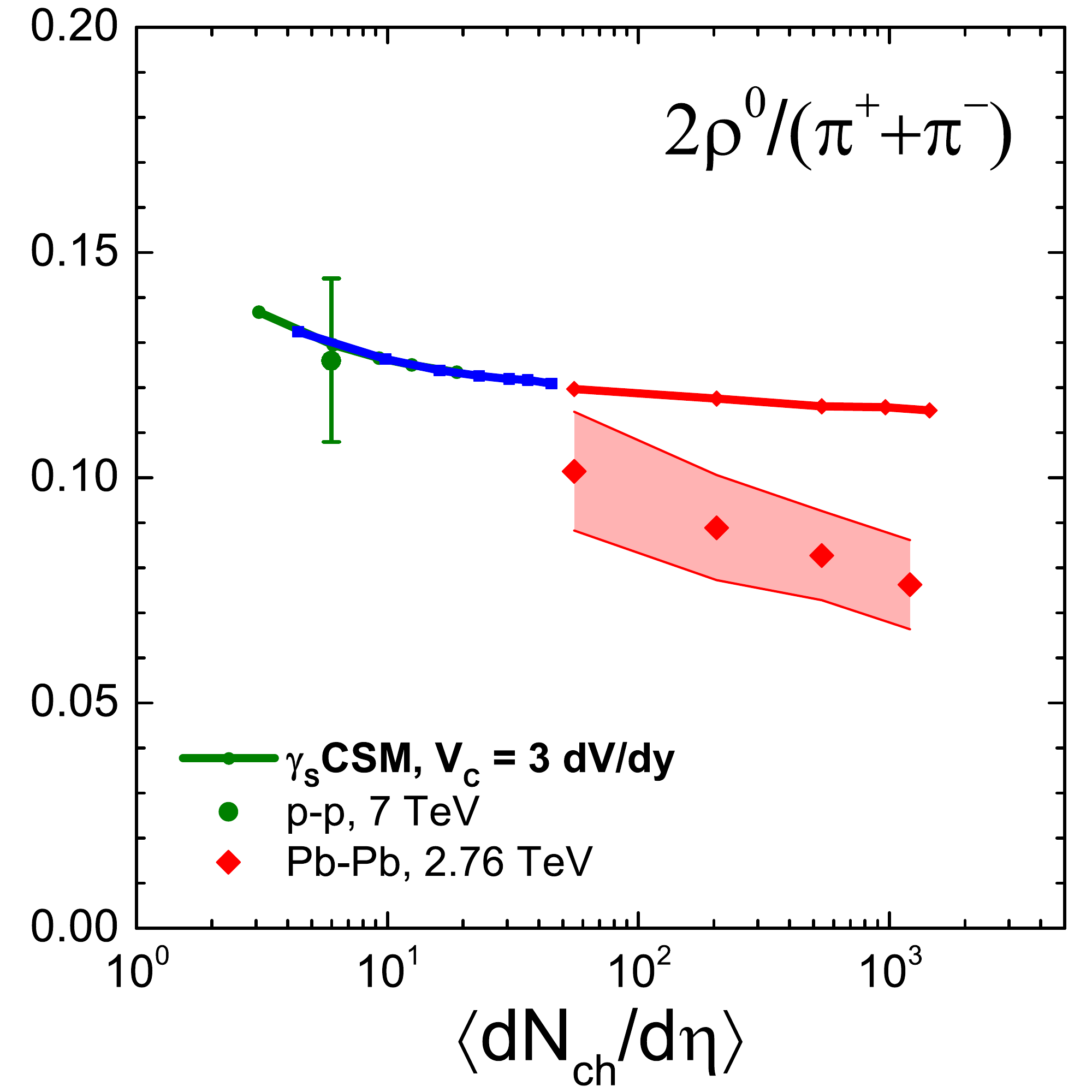}
  \includegraphics[width=.32\textwidth]{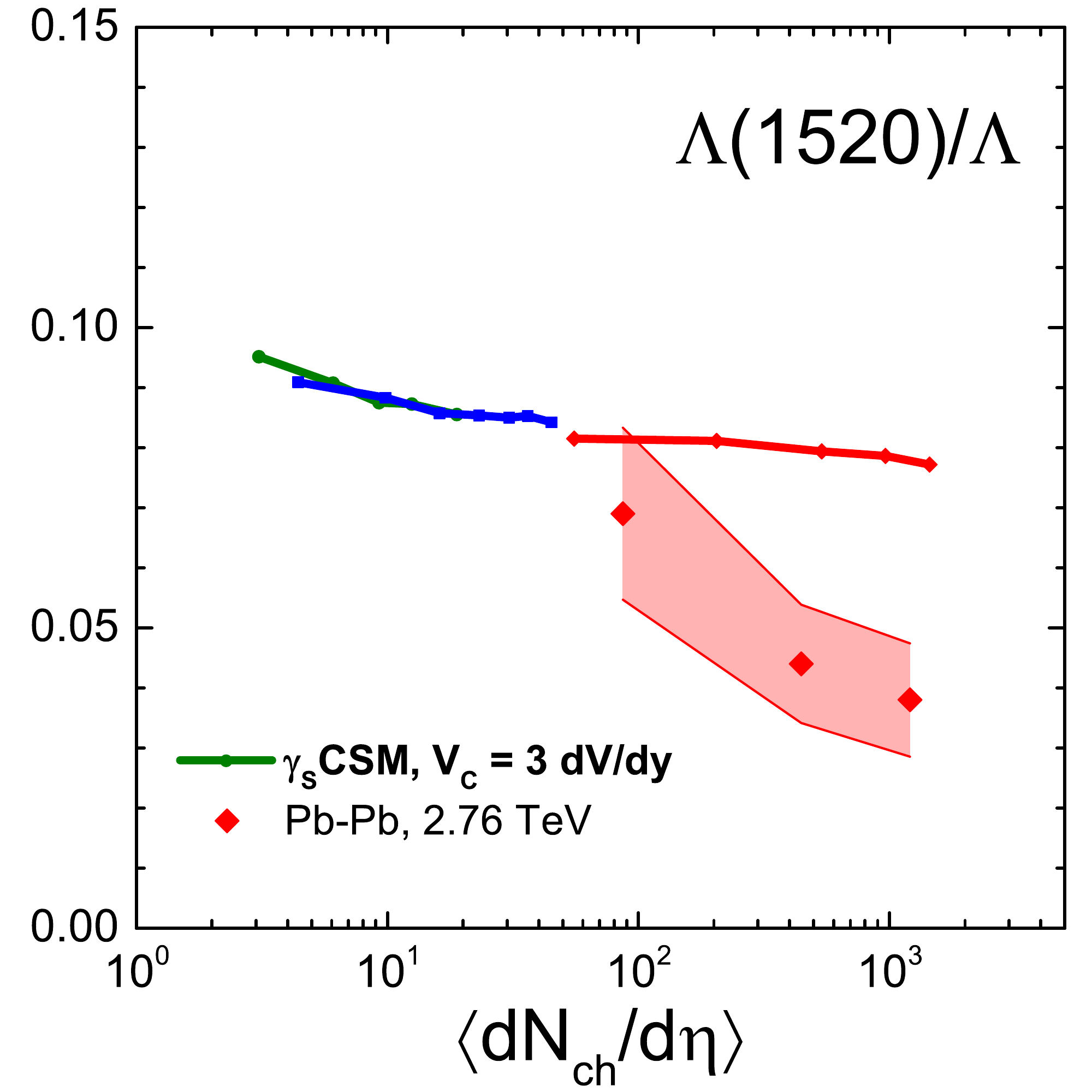}
  \includegraphics[width=.32\textwidth]{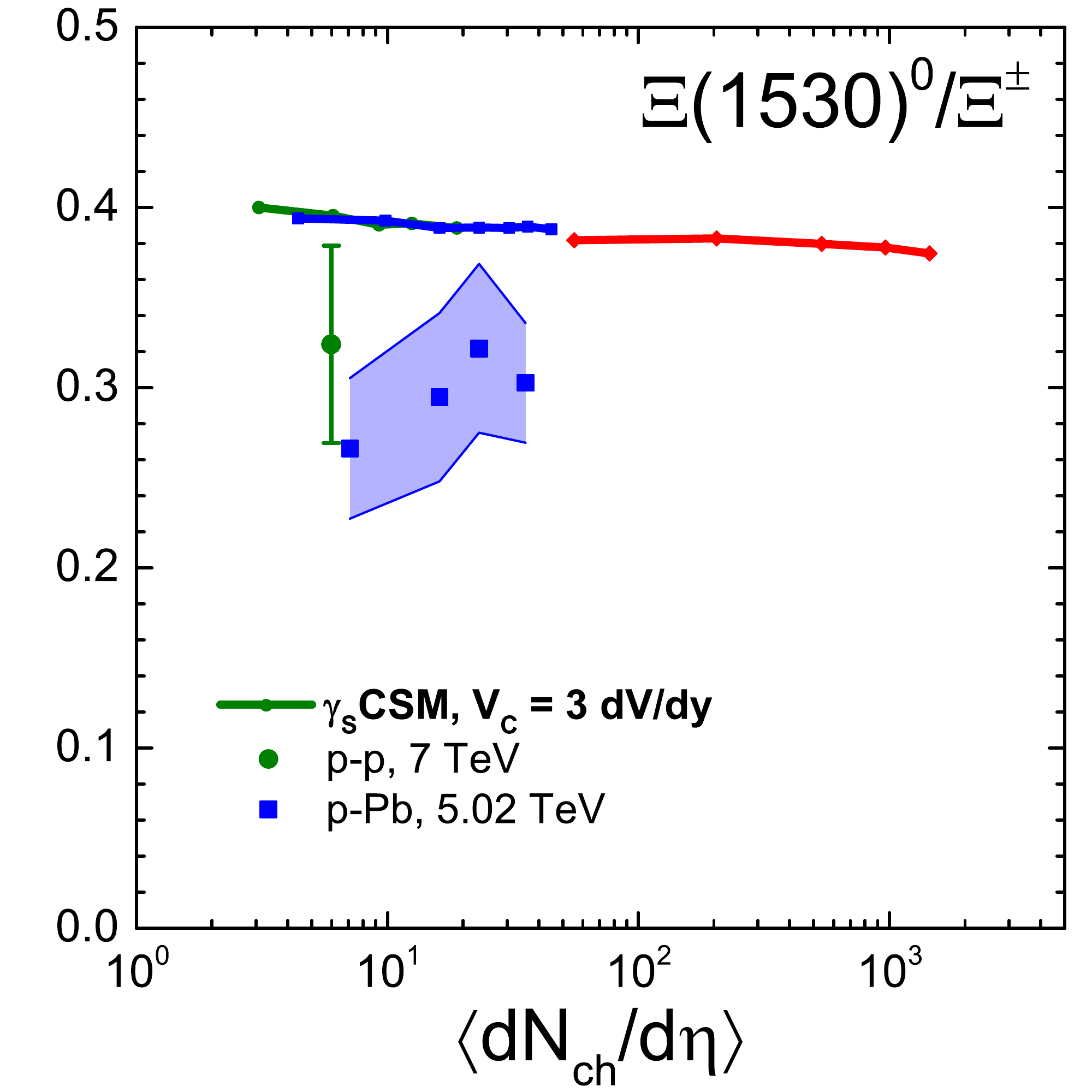}
  \includegraphics[width=.32\textwidth]{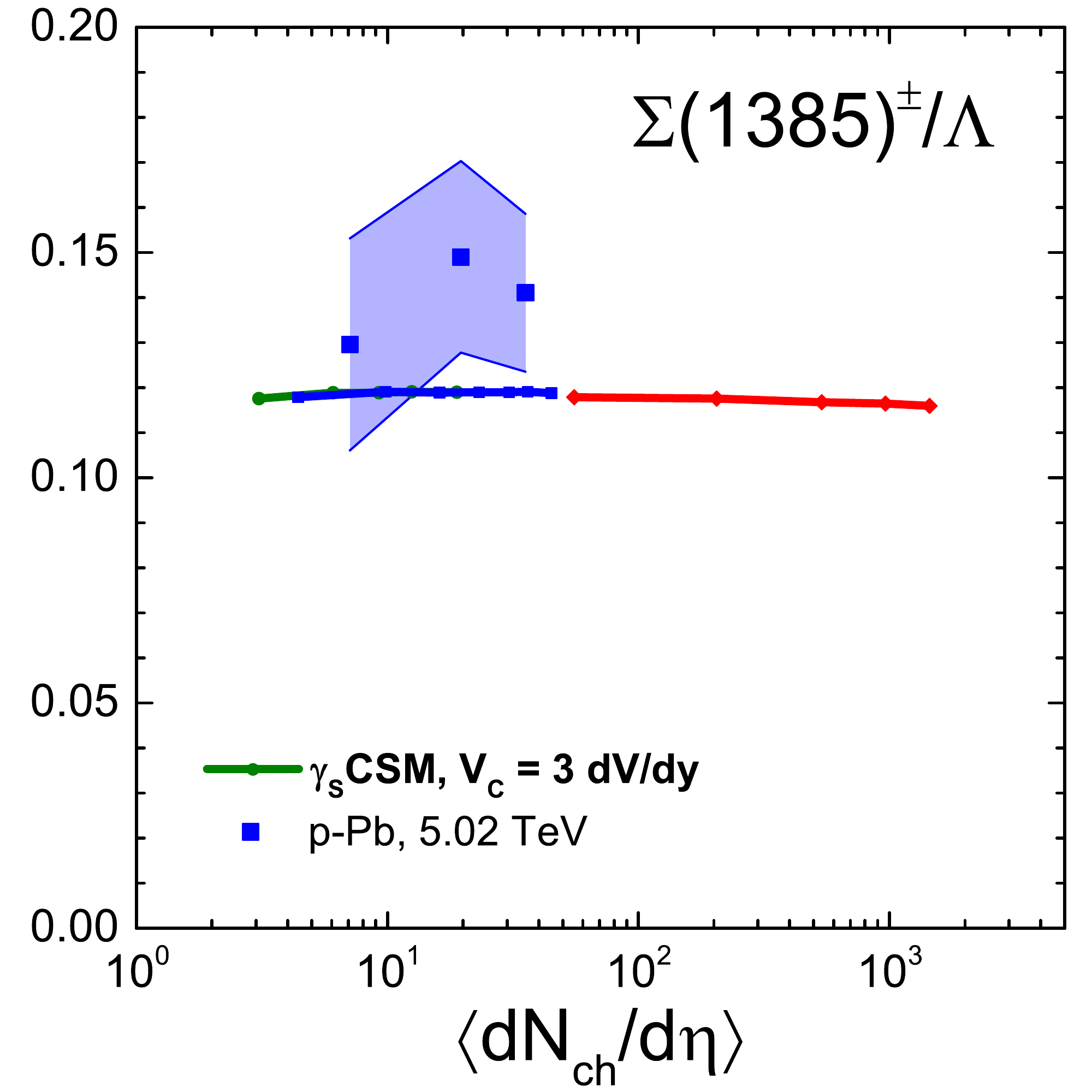}
  \includegraphics[width=.32\textwidth]{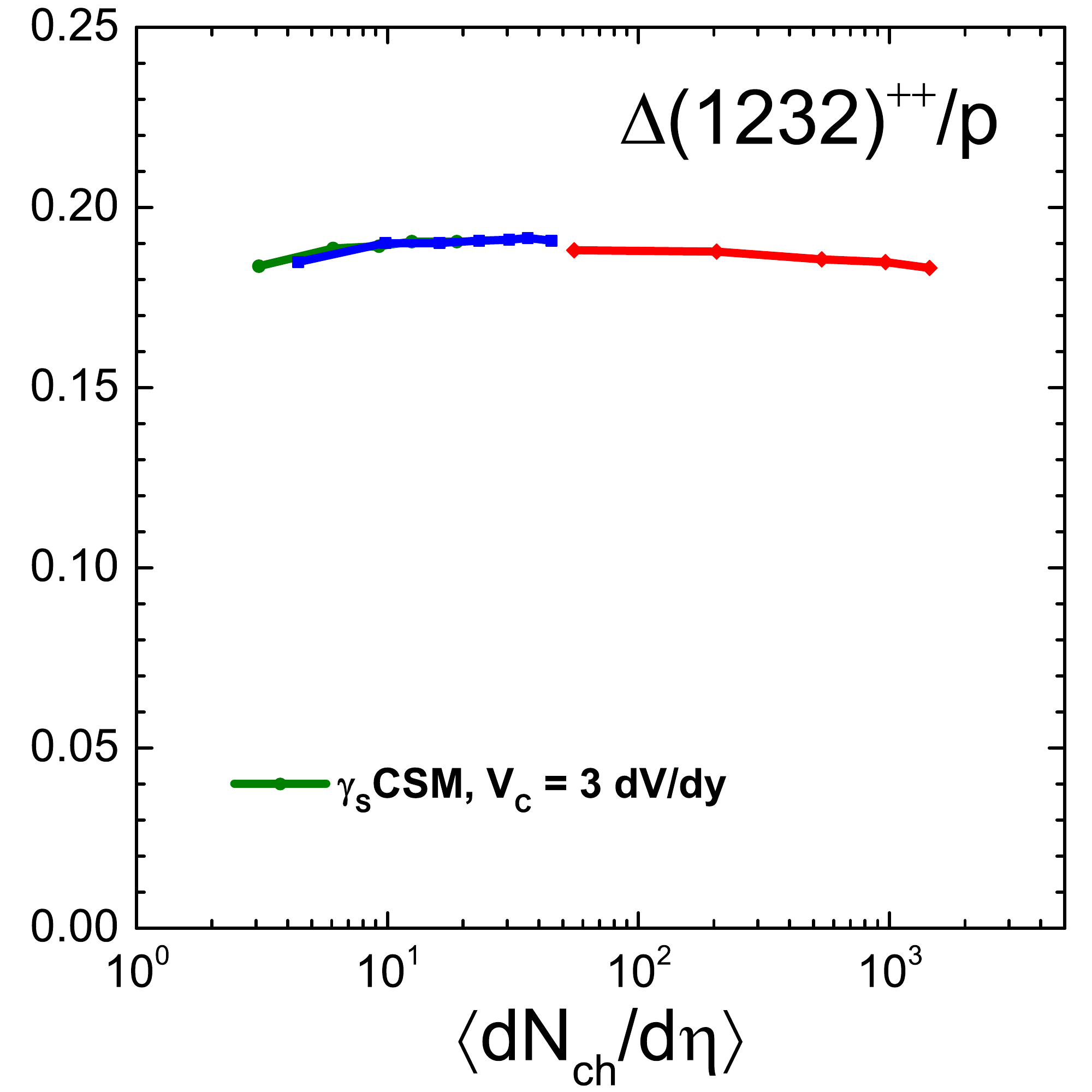}
  \caption{
  The dependence of yield ratios (a) $K^*/K$, (b) $2\rho^0/(\pi^+ + \pi^-)$, (c) $\Lambda(1520) / \Lambda$, (d) $\Xi(1530)^0/\Xi^{\pm}$, (e) $\Sigma^\pm(1385)/\Lambda$, and (f) $\Delta(1232)^{++}$/p evaluated in the $\gamma_S$CSM with $V_c = 3 \, dV/dy$ for the thermal parameters extracted for each multiplicity bin through the $\chi^2$ minimization of yields of stable hadrons.
  The green circles, blue squares, and red diamonds depict the corresponding ratios measured by the ALICE collaboration in p-p~(7~TeV)~\cite{Abelev:2014qqa,Acharya:2018orn,Acharya:2018qnp}, p-Pb~(5.02~TeV)~\cite{Adam:2016bpr,Adamova:2017elh}, and Pb-Pb~(2.76~TeV)~\cite{Abelev:2014uua,Acharya:2018qnp,ALICE:2018ewo} collisions where available.
  }
  \label{fig:CSMResonances}
\end{figure*}

Figure~\ref{fig:CSMFittedYields} shows the dependence of the yield ratios $p/\pi$, $K/\pi$, $\phi / \pi$, $\Lambda / \pi$, $\Xi/\pi$, and $\Omega / \pi$ on the charged particle multiplicity $\langle d N_{\rm ch} / d \eta \rangle_{|\eta| < 0.5}$, as evaluated in the $\gamma_S$CSM with $V_c = 3 \, dV/dy$ for the thermal parameters extracted from the fits.
The $\gamma_S$CSM reproduces quite well the trends observed in the data.
This also includes the rather abrupt jump in the $\Xi/\pi$ ratio when going from peripheral Pb--Pb collisions to most central p--Pb collisions.
Obviously, this is a result of fitting the model to data independently in each multiplicity bin.
It should be noted that the $\Xi$ yield data in Pb--Pb collisions are currently under re-analysis~\cite{Kalinak,Kalinak:2017xll} within the ALICE collaboration, and a re-fitting of the $\gamma_S$CSM might be required if corrected data become available.

The model overpredicts systematically the $p/\pi$ ratio, roughly on a $2\sigma$ level.
Separately, the proton yields are overpredicted on a $1\sigma$ level, while the yields of pions are underpredicted on a $1\sigma$ level.
Note that here energy-dependent Breit-Wigner widths are used, which reduce the $p/\pi$ ratios by about 15\% relative to the common zero-width approximation~\cite{Vovchenko:2018fmh}.
If the zero-width approximation would be used instead, then the tension with the $p/\pi$ data would be even larger.
The application of energy-dependent Breit-Wigner widths leads to a systematic improvement of the data description at \emph{all} multiplicities, although the description of the $p/\pi$ ratio is still not fully satisfactory.

The temperature and $\gamma_S$ values extracted for the most central Pb--Pb bin are consistent, within errors, with the $T_{\rm ch} = 155$~MeV and $\gamma_S = 1$ values of the vanilla CSM in Sec.~\ref{sec:VanillaCSM}.
The data description accuracy between Figs.~\ref{fig:CSMYieldRatios} and~\ref{fig:CSMFittedYields} is almost identical for the highest multiplicity bin.
At lower multiplicities, however, differences between vanilla CSM and $\gamma_S$CSM become more and more significant.

Model predictions can further be cross-checked with the data on those hadron yields which are not used in the fit procedure. 
The yields of resonances are particularly interesting in this regard:
the yield ratios $K^{*0}/K^-$, $2\rho^0 / (\pi^+ + \pi^-)$, and $\Lambda(1520)/\Lambda$ have been measured by the ALICE collaboration in p--p, p--Pb, and Pb--Pb collisions~\cite{Abelev:2014uua,Acharya:2018qnp,ALICE:2018ewo}.
The effects associated with exact conservation of baryon number and strangeness, as well incomplete equilibration of strangeness, do in fact cancel out in these ratios to the leading order.
These ratios, however, are potentially sensitive to changes of the freeze-out temperature.
These ratios are depicted in Fig.~\ref{fig:CSMResonances} as a function of the charged multiplicity $\langle d N_{\rm ch} / d \eta \rangle_{|\eta| < 0.5}$, evaluated in the $\gamma_S$CSM with $V_c = 3 \, dV/dy$ for the thermal parameters extracted from the fits.
The increase in the chemical freeze-out temperature for lower multiplicities, seen in Fig.~\ref{fig:CSMFitParams}, leads to an increase of the above-mentioned ratios at small multiplicities.
This increase is moderate, and within the experimental error bars. 
The effect is more moderate than can be expected based on simple considerations regarding the Boltzmann factor, the mass difference between resonances and corresponding stable particles, and a change in the freeze-out temperature depending on the multiplicity. 
The reason for that are the significant feeddown contributions to both the numerator and denumerator, which weaken the influence of the changing temperature.

The available data for p--p and p--Pb collisions is described fairly well.
Resonance yields are overestimated in central Pb--Pb collisions.
This suppression of the measured yields is often interpreted as an evidence for a hadronic phase after the chemical freeze-out~\cite{Abelev:2014uua,Acharya:2018qnp,ALICE:2018ewo}.
In this case the thermal picture should be extended to incorporate the hadronic phase, for instance using the concept of partial chemical equilibrium~\cite{Bebie:1991ij} or a hadronic afterburner. Both cases lead to suppressed yields of short-lived resonances relative to the chemical equilibrium statistical model predictions~\cite{Vovchenko:2019aoz,Steinheimer:2017vju}.

We also present in Fig.~\ref{fig:CSMResonances} predictions for the yield ratios $\Xi(1530)^0/\Xi^{\pm}$, $\Sigma^\pm(1385)/\Lambda$, and $\Sigma^-(1385)/\Lambda$, along with the available p--p~\cite{Abelev:2014qqa} and p--Pb~\cite{Adamova:2017elh} data. Among these, the $\Xi(1530)^0/\Xi^{\pm}$ ratio does show a mild multiplicity dependence, with the smallest values at the highest multiplicities. The multiplicity dependence of $\Sigma^-(1385)/\Lambda$ and $\Delta^{++}(1232)$/p is more moderate.
A comparison with prospective experimental measurements  of these ratios in Pb--Pb collisions would be important, as it will be particularly interesting to see whether the data will show a suppression at large multiplicities relative to the ($\gamma_S$)CSM predictions.

\subsection{Quantifying the data description accuracy}

\begin{figure*}[!ht]
  \centering
  \includegraphics[width=.99\textwidth]{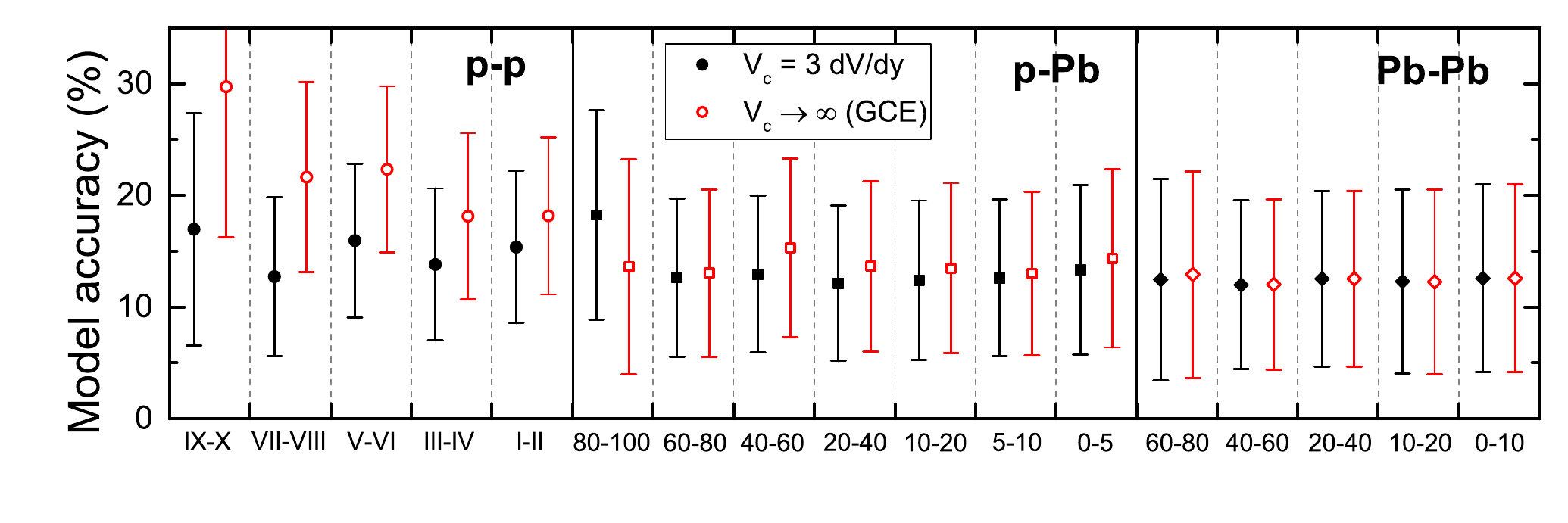}
  \caption{
  The relative accuracy of p--p~(circles), p--Pb~(squares), and Pb--Pb~(diamonds) ALICE data description within the $\gamma_S$CSM with $V_c = 3 \, dV/dy$~(full black symbols) and in the grand-canonical limit $V_c \to \infty$~(open red symbols).
  }
  \label{fig:CSMAccuracy}
\end{figure*}

The $\chi^2$ values extracted from the fits are 2-3 times larger for p--p and for p--Pb collisions as compared to those for Pb--Pb.
This might indicate a better performance of the $\gamma_S$CSM for large multiplicities compared to the lower ones.
We argue, however, that these $\chi^2$ values rather reflect the differences in the measurement uncertainties between the different colliding systems analyzed rather than the performance of the model.

The simple statistical model has only a certain relative accuracy in describing the yield of hadrons produced in heavy-ion collisions.
Smaller measurement uncertainties will inevitably lead to increased values of $\chi^2$ at some point, without necessarily implying that a relative accuracy of the thermal model has worsened.
In order to quantify the relative model accuracy we introduce the following measure of describing the hadron yield data:
\eq{\label{eq:accuracy}
\kappa = \sum_i w_i \, \left| \frac{\langle N_i^{\rm mod} \rangle}{\langle N_i^{\rm exp} \rangle} - 1 \right|~.
}
Here the sum goes over all species in a thermal fit. 
The weights $w_i$ are proportional to the contribution of the data point $i$ to $\chi^2$:
\eq{
w_i = \frac{(\langle N_i^{\rm mod} \rangle - \langle N_i^{\rm exp} \rangle)^2 / \sigma_i^2}{\chi^2},
}
with $\sigma_i$ being the measurement uncertainty for the hadron yield $i$.
The weights evidently satisfy the normalization condition $\sum_i w_i = 1$.
The uncertainty of the $\kappa$ value is estimated as follows:
\eq{
\delta \kappa = \sum_i w_i \, \frac{\langle N_i^{\rm mod} \rangle \, \sigma_i}{(\langle N_i^{\rm exp} \rangle)^2}~.
}

$\kappa$ quantifies the average relative accuracy with which the model describes the given data set.
For example, if a model describes the data on a 10-15\% level accuracy, the calculated value of $\kappa$ ought to lie within 0.10-0.15.

Figure~\ref{fig:CSMAccuracy} depicts the computed values of $\kappa$ for the data from p--p, p--Pb, and Pb--Pb collisions at different centralities for the $\gamma_S$CSM with $V_c = 3 \, dV/dy$~(full black symbols), and the $\gamma_S$CSM in the grand-canonical limit~(open red symbols).
The $\gamma_S$CSM with $V_c = 3 \, dV/dy$ describes the data roughly 15\% on a level according to this newly introduced measure, uniformly across \emph{all} multiplicities, Fig.~\ref{fig:CSMAccuracy}.
The grand-canonical version of the statistical model, on the other hand, illustrates the trend that the model accuracy gets worse as one goes from Pb--Pb/p--Pb collisions to p--p collisions, especially for the smallest three p--p multiplicity bins.

\section{Summary}

We analyzed the multiplicity dependence of the hadron yields measured by the ALICE collaboration at the LHC within the statistical model with exact conservation of baryon number, electric charge, and strangeness.
We find that the conservation of baryon number is at least as important as the exact strangeness conservation in the canonical-statistical picture at the LHC. This is in stark contrast to low and intermediate collision energies~($\sqrt{s_{\rm NN}} \lesssim 10$~GeV), where strangeness-canonical ensemble is sufficient for many applications where the canonical suppression is important.

The effects of exact conservation of conserved charges leads to suppression of yields of most hadron species relative to their grand-canonical values, with a notable exception of $\phi$ mesons, their yields being unaffected by canonical suppression.
The grand canonical ensemble statistical model does not only theoretically underpredict the hadron yield data in small systems at the LHC, but it is shown that only a combined canonical treatment of all three conserved charges yields a reasonable description.

The simplified version of the CSM~(the vanilla CSM) assumes a constant freeze-out temperature of $T \simeq 155$~MeV across all multiplicity bins, and the multiplicity dependence of hadron yield ratios is driven by their dependence on the correlation volume $V_c$ only. The vanilla CSM captures fairly well multiplicity dependence of hyperon-to-pion and light nuclei-to-proton yield ratios. 
The trend in the $K/\pi$ ratio is also captured, but the model overshoots the data significantly, except for the most central Pb--Pb collisions.
The $p/\pi$ ratio is affected by canonical suppression in the vanilla CSM. The data, on the other hand, show no clear evidence for the canonical suppression. 
The model does not describe the proton-to-pion and hyperon-to-pion ratios simultaneously: a large correlation volume over 5-6 units of rapidity is needed to accommodate the former ratio whereas a smaller (around 3 rapidity units) volume is needed for the latter. 
The behavior of the $\phi/\pi$ ratio in the vanilla CSM is opposite to the behavior in the data. Thus, unless production mechanism of $\phi$'s separate from the rest of hadrons, this invalidates the vanilla CSM for p-p and p-Pb collisions.

In a more complete $\gamma_S$CSM study we consider a multiplicity-dependent chemical freeze-out temperature, including a possibility of an incomplete chemical equilibration in the strangeness sector, and assume that the canonical correlation volume corresponds to three units of rapidity, i.e. $V_c = 3 \, dV/dy$.
The latter assumption is in a fair agreement with the available preliminary data on net-proton fluctuations in  Pb--Pb collisions at the LHC~\cite{Rustamov:2017lio}~(Sec.~\ref{sec:flucs}).
Fits to the ALICE data for various multiplicity bins in p--p, p--Pb, and Pb--Pb collisions indicate the preference of the canonical approach over the grand-canonical one for p--p collisions. This is not the case for p--Pb and Pb--Pb collisions. Apparent reasons for that are $\Omega$'s, which are measured with a better precision in p-p, and the fact that the canonical suppression is partially mimicked in the grand-canonical picture through smaller values of $\gamma_S$.

The chemical freeze-out temperature extracted with the $\gamma_S$CSM is found to decrease with increasing charged particle multiplicity. 
It reaches a maximum value of $T \simeq 175$~MeV for the smallest multiplicity bins in p--p and the minimum value of $T \simeq 155$~MeV for the highest multiplicities in Pb--Pb.
Higher extracted temperatures for smaller systems might indicate an earlier chemical freeze-out in those systems, similar to an observed earlier kinetic freeze-out in smaller systems within the blast-wave picture.
It is also notable that the extracted temperatures for the smallest systems are larger than the lattice QCD estimates for the pseudocritical temperature $T_{\rm pc} = 155$~MeV of the chiral crossover transition in the infinite volume limit~\cite{Bazavov:2018mes}.
The strangeness saturation parameter $\gamma_S$ increases with multiplicity, reaching the limiting value of unity at $d N_{\rm ch} / d \eta \simeq 100-200$, indicating that chemical equilibrium is  established only in sufficiently large systems.
The canonical suppression effects for hadron yields are found to become virtually negligible for large enough multiplicities, $d N_{\rm ch} / d \eta \gtrsim 50$.

Analysis of experimental data indicates that a statistical model approach with exact conservation of baryon number, electric charge, and strangeness, and an incomplete chemical equilibration of strangeness is capable of describing the hadron yields measured by the ALICE collaboration with roughly a 15\% relative accuracy, across all multiplicity bins measured so far.
The canonical suppression as well as effects of multiplicity-dependent freeze-out temperature do not significantly alter the systematics of resonance production.
Improved measurements across a large range of multiplicities coming from the LHC Run 2 will shed further light on the (canonical-)thermal aspects of particle production in systems of various sizes.\\
\,
\vskip2cm

\begin{acknowledgments}

We thank J\"urgen Schukraft for motivating discussions and a suggestion to consider the yields of resonances.
B.D. acknowledges the support from BMBF through the FSP202 (F\"orderkennzeichen 05P15RFCA1).
H.St. acknowledges the support through the Judah M. Eisenberg Laureatus Chair by Goethe University  and the Walter Greiner Gesellschaft, Frankfurt.

\end{acknowledgments}


\bibliography{CSM-LHC}

\begin{thebibliography}{58}%
\makeatletter
\providecommand \@ifxundefined [1]{%
 \@ifx{#1\undefined}
}%
\providecommand \@ifnum [1]{%
 \ifnum #1\expandafter \@firstoftwo
 \else \expandafter \@secondoftwo
 \fi
}%
\providecommand \@ifx [1]{%
 \ifx #1\expandafter \@firstoftwo
 \else \expandafter \@secondoftwo
 \fi
}%
\providecommand \natexlab [1]{#1}%
\providecommand \enquote  [1]{``#1''}%
\providecommand \bibnamefont  [1]{#1}%
\providecommand \bibfnamefont [1]{#1}%
\providecommand \citenamefont [1]{#1}%
\providecommand \href@noop [0]{\@secondoftwo}%
\providecommand \href [0]{\begingroup \@sanitize@url \@href}%
\providecommand \@href[1]{\@@startlink{#1}\@@href}%
\providecommand \@@href[1]{\endgroup#1\@@endlink}%
\providecommand \@sanitize@url [0]{\catcode `\\12\catcode `\$12\catcode
  `\&12\catcode `\#12\catcode `\^12\catcode `\_12\catcode `\%12\relax}%
\providecommand \@@startlink[1]{}%
\providecommand \@@endlink[0]{}%
\providecommand \url  [0]{\begingroup\@sanitize@url \@url }%
\providecommand \@url [1]{\endgroup\@href {#1}{\urlprefix }}%
\providecommand \urlprefix  [0]{URL }%
\providecommand \Eprint [0]{\href }%
\providecommand \doibase [0]{http://dx.doi.org/}%
\providecommand \selectlanguage [0]{\@gobble}%
\providecommand \bibinfo  [0]{\@secondoftwo}%
\providecommand \bibfield  [0]{\@secondoftwo}%
\providecommand \translation [1]{[#1]}%
\providecommand \BibitemOpen [0]{}%
\providecommand \bibitemStop [0]{}%
\providecommand \bibitemNoStop [0]{.\EOS\space}%
\providecommand \EOS [0]{\spacefactor3000\relax}%
\providecommand \BibitemShut  [1]{\csname bibitem#1\endcsname}%
\let\auto@bib@innerbib\@empty
\bibitem [{\citenamefont {Acharya}\ \emph
  {et~al.}(2019{\natexlab{a}})\citenamefont {Acharya} \emph
  {et~al.}}]{Acharya:2018orn}%
  \BibitemOpen
  \bibfield  {author} {\bibinfo {author} {\bibfnamefont {S.}~\bibnamefont
  {Acharya}} \emph {et~al.} (\bibinfo {collaboration} {ALICE}),\ }\href
  {\doibase 10.1103/PhysRevC.99.024906} {\bibfield  {journal} {\bibinfo
  {journal} {Phys. Rev.}\ }\textbf {\bibinfo {volume} {C99}},\ \bibinfo {pages}
  {024906} (\bibinfo {year} {2019}{\natexlab{a}})},\ \Eprint
  {http://arxiv.org/abs/1807.11321} {arXiv:1807.11321 [nucl-ex]} \BibitemShut
  {NoStop}%
\bibitem [{\citenamefont {Abelev}\ \emph
  {et~al.}(2014{\natexlab{a}})\citenamefont {Abelev} \emph
  {et~al.}}]{Abelev:2013haa}%
  \BibitemOpen
  \bibfield  {author} {\bibinfo {author} {\bibfnamefont {B.~B.}\ \bibnamefont
  {Abelev}} \emph {et~al.} (\bibinfo {collaboration} {ALICE}),\ }\href
  {\doibase 10.1016/j.physletb.2013.11.020} {\bibfield  {journal} {\bibinfo
  {journal} {Phys. Lett.}\ }\textbf {\bibinfo {volume} {B728}},\ \bibinfo
  {pages} {25} (\bibinfo {year} {2014}{\natexlab{a}})},\ \Eprint
  {http://arxiv.org/abs/1307.6796} {arXiv:1307.6796 [nucl-ex]} \BibitemShut
  {NoStop}%
\bibitem [{\citenamefont {Adam}\ \emph
  {et~al.}(2016{\natexlab{a}})\citenamefont {Adam} \emph
  {et~al.}}]{Adam:2015vsf}%
  \BibitemOpen
  \bibfield  {author} {\bibinfo {author} {\bibfnamefont {J.}~\bibnamefont
  {Adam}} \emph {et~al.} (\bibinfo {collaboration} {ALICE}),\ }\href {\doibase
  10.1016/j.physletb.2016.05.027} {\bibfield  {journal} {\bibinfo  {journal}
  {Phys. Lett.}\ }\textbf {\bibinfo {volume} {B758}},\ \bibinfo {pages} {389}
  (\bibinfo {year} {2016}{\natexlab{a}})},\ \Eprint
  {http://arxiv.org/abs/1512.07227} {arXiv:1512.07227 [nucl-ex]} \BibitemShut
  {NoStop}%
\bibitem [{\citenamefont {Adam}\ \emph
  {et~al.}(2016{\natexlab{b}})\citenamefont {Adam} \emph
  {et~al.}}]{Adam:2016bpr}%
  \BibitemOpen
  \bibfield  {author} {\bibinfo {author} {\bibfnamefont {J.}~\bibnamefont
  {Adam}} \emph {et~al.} (\bibinfo {collaboration} {ALICE}),\ }\href {\doibase
  10.1140/epjc/s10052-016-4088-7} {\bibfield  {journal} {\bibinfo  {journal}
  {Eur. Phys. J.}\ }\textbf {\bibinfo {volume} {C76}},\ \bibinfo {pages} {245}
  (\bibinfo {year} {2016}{\natexlab{b}})},\ \Eprint
  {http://arxiv.org/abs/1601.07868} {arXiv:1601.07868 [nucl-ex]} \BibitemShut
  {NoStop}%
\bibitem [{\citenamefont {Abelev}\ \emph
  {et~al.}(2013{\natexlab{a}})\citenamefont {Abelev} \emph
  {et~al.}}]{Abelev:2013vea}%
  \BibitemOpen
  \bibfield  {author} {\bibinfo {author} {\bibfnamefont {B.}~\bibnamefont
  {Abelev}} \emph {et~al.} (\bibinfo {collaboration} {ALICE}),\ }\href
  {\doibase 10.1103/PhysRevC.88.044910} {\bibfield  {journal} {\bibinfo
  {journal} {Phys. Rev.}\ }\textbf {\bibinfo {volume} {C88}},\ \bibinfo {pages}
  {044910} (\bibinfo {year} {2013}{\natexlab{a}})},\ \Eprint
  {http://arxiv.org/abs/1303.0737} {arXiv:1303.0737 [hep-ex]} \BibitemShut
  {NoStop}%
\bibitem [{\citenamefont {Abelev}\ \emph
  {et~al.}(2013{\natexlab{b}})\citenamefont {Abelev} \emph
  {et~al.}}]{Abelev:2013xaa}%
  \BibitemOpen
  \bibfield  {author} {\bibinfo {author} {\bibfnamefont {B.~B.}\ \bibnamefont
  {Abelev}} \emph {et~al.} (\bibinfo {collaboration} {ALICE}),\ }\href
  {\doibase 10.1103/PhysRevLett.111.222301} {\bibfield  {journal} {\bibinfo
  {journal} {Phys. Rev. Lett.}\ }\textbf {\bibinfo {volume} {111}},\ \bibinfo
  {pages} {222301} (\bibinfo {year} {2013}{\natexlab{b}})},\ \Eprint
  {http://arxiv.org/abs/1307.5530} {arXiv:1307.5530 [nucl-ex]} \BibitemShut
  {NoStop}%
\bibitem [{\citenamefont {Abelev}\ \emph
  {et~al.}(2014{\natexlab{b}})\citenamefont {Abelev} \emph
  {et~al.}}]{ABELEV:2013zaa}%
  \BibitemOpen
  \bibfield  {author} {\bibinfo {author} {\bibfnamefont {B.~B.}\ \bibnamefont
  {Abelev}} \emph {et~al.} (\bibinfo {collaboration} {ALICE}),\ }\href
  {\doibase 10.1016/j.physletb.2014.05.052, 10.1016/j.physletb.2013.11.048}
  {\bibfield  {journal} {\bibinfo  {journal} {Phys. Lett.}\ }\textbf {\bibinfo
  {volume} {B728}},\ \bibinfo {pages} {216} (\bibinfo {year}
  {2014}{\natexlab{b}})},\ \bibinfo {note} {[Erratum: Phys.
  Lett.B734,409(2014)]},\ \Eprint {http://arxiv.org/abs/1307.5543}
  {arXiv:1307.5543 [nucl-ex]} \BibitemShut {NoStop}%
\bibitem [{\citenamefont {Abelev}\ \emph
  {et~al.}(2015{\natexlab{a}})\citenamefont {Abelev} \emph
  {et~al.}}]{Abelev:2014uua}%
  \BibitemOpen
  \bibfield  {author} {\bibinfo {author} {\bibfnamefont {B.~B.}\ \bibnamefont
  {Abelev}} \emph {et~al.} (\bibinfo {collaboration} {ALICE}),\ }\href
  {\doibase 10.1103/PhysRevC.91.024609} {\bibfield  {journal} {\bibinfo
  {journal} {Phys. Rev.}\ }\textbf {\bibinfo {volume} {C91}},\ \bibinfo {pages}
  {024609} (\bibinfo {year} {2015}{\natexlab{a}})},\ \Eprint
  {http://arxiv.org/abs/1404.0495} {arXiv:1404.0495 [nucl-ex]} \BibitemShut
  {NoStop}%
\bibitem [{\citenamefont {Sjostrand}\ \emph {et~al.}(2008)\citenamefont
  {Sjostrand}, \citenamefont {Mrenna},\ and\ \citenamefont
  {Skands}}]{Sjostrand:2007gs}%
  \BibitemOpen
  \bibfield  {author} {\bibinfo {author} {\bibfnamefont {T.}~\bibnamefont
  {Sjostrand}}, \bibinfo {author} {\bibfnamefont {S.}~\bibnamefont {Mrenna}}, \
  and\ \bibinfo {author} {\bibfnamefont {P.~Z.}\ \bibnamefont {Skands}},\
  }\href {\doibase 10.1016/j.cpc.2008.01.036} {\bibfield  {journal} {\bibinfo
  {journal} {Comput. Phys. Commun.}\ }\textbf {\bibinfo {volume} {178}},\
  \bibinfo {pages} {852} (\bibinfo {year} {2008})},\ \Eprint
  {http://arxiv.org/abs/0710.3820} {arXiv:0710.3820 [hep-ph]} \BibitemShut
  {NoStop}%
\bibitem [{\citenamefont {Skands}\ \emph {et~al.}(2014)\citenamefont {Skands},
  \citenamefont {Carrazza},\ and\ \citenamefont {Rojo}}]{Skands:2014pea}%
  \BibitemOpen
  \bibfield  {author} {\bibinfo {author} {\bibfnamefont {P.}~\bibnamefont
  {Skands}}, \bibinfo {author} {\bibfnamefont {S.}~\bibnamefont {Carrazza}}, \
  and\ \bibinfo {author} {\bibfnamefont {J.}~\bibnamefont {Rojo}},\ }\href
  {\doibase 10.1140/epjc/s10052-014-3024-y} {\bibfield  {journal} {\bibinfo
  {journal} {Eur. Phys. J.}\ }\textbf {\bibinfo {volume} {C74}},\ \bibinfo
  {pages} {3024} (\bibinfo {year} {2014})},\ \Eprint
  {http://arxiv.org/abs/1404.5630} {arXiv:1404.5630 [hep-ph]} \BibitemShut
  {NoStop}%
\bibitem [{\citenamefont {Bierlich}\ \emph {et~al.}(2015)\citenamefont
  {Bierlich}, \citenamefont {Gustafson}, \citenamefont {Lönnblad},\ and\
  \citenamefont {Tarasov}}]{Bierlich:2014xba}%
  \BibitemOpen
  \bibfield  {author} {\bibinfo {author} {\bibfnamefont {C.}~\bibnamefont
  {Bierlich}}, \bibinfo {author} {\bibfnamefont {G.}~\bibnamefont {Gustafson}},
  \bibinfo {author} {\bibfnamefont {L.}~\bibnamefont {Lönnblad}}, \ and\
  \bibinfo {author} {\bibfnamefont {A.}~\bibnamefont {Tarasov}},\ }\href
  {\doibase 10.1007/JHEP03(2015)148} {\bibfield  {journal} {\bibinfo  {journal}
  {JHEP}\ }\textbf {\bibinfo {volume} {03}},\ \bibinfo {pages} {148} (\bibinfo
  {year} {2015})},\ \Eprint {http://arxiv.org/abs/1412.6259} {arXiv:1412.6259
  [hep-ph]} \BibitemShut {NoStop}%
\bibitem [{\citenamefont {Pierog}\ \emph {et~al.}(2015)\citenamefont {Pierog},
  \citenamefont {Karpenko}, \citenamefont {Katzy}, \citenamefont {Yatsenko},\
  and\ \citenamefont {Werner}}]{Pierog:2013ria}%
  \BibitemOpen
  \bibfield  {author} {\bibinfo {author} {\bibfnamefont {T.}~\bibnamefont
  {Pierog}}, \bibinfo {author} {\bibfnamefont {I.}~\bibnamefont {Karpenko}},
  \bibinfo {author} {\bibfnamefont {J.~M.}\ \bibnamefont {Katzy}}, \bibinfo
  {author} {\bibfnamefont {E.}~\bibnamefont {Yatsenko}}, \ and\ \bibinfo
  {author} {\bibfnamefont {K.}~\bibnamefont {Werner}},\ }\href {\doibase
  10.1103/PhysRevC.92.034906} {\bibfield  {journal} {\bibinfo  {journal} {Phys.
  Rev.}\ }\textbf {\bibinfo {volume} {C92}},\ \bibinfo {pages} {034906}
  (\bibinfo {year} {2015})},\ \Eprint {http://arxiv.org/abs/1306.0121}
  {arXiv:1306.0121 [hep-ph]} \BibitemShut {NoStop}%
\bibitem [{\citenamefont {Petrán}\ \emph {et~al.}(2013)\citenamefont
  {Petrán}, \citenamefont {Letessier}, \citenamefont {Petráček},\ and\
  \citenamefont {Rafelski}}]{Petran:2013lja}%
  \BibitemOpen
  \bibfield  {author} {\bibinfo {author} {\bibfnamefont {M.}~\bibnamefont
  {Petrán}}, \bibinfo {author} {\bibfnamefont {J.}~\bibnamefont {Letessier}},
  \bibinfo {author} {\bibfnamefont {V.}~\bibnamefont {Petráček}}, \ and\
  \bibinfo {author} {\bibfnamefont {J.}~\bibnamefont {Rafelski}},\ }\href
  {\doibase 10.1103/PhysRevC.88.034907} {\bibfield  {journal} {\bibinfo
  {journal} {Phys. Rev.}\ }\textbf {\bibinfo {volume} {C88}},\ \bibinfo {pages}
  {034907} (\bibinfo {year} {2013})},\ \Eprint {http://arxiv.org/abs/1303.2098}
  {arXiv:1303.2098 [hep-ph]} \BibitemShut {NoStop}%
\bibitem [{\citenamefont {Stachel}\ \emph {et~al.}(2014)\citenamefont
  {Stachel}, \citenamefont {Andronic}, \citenamefont {Braun-Munzinger},\ and\
  \citenamefont {Redlich}}]{Stachel:2013zma}%
  \BibitemOpen
  \bibfield  {author} {\bibinfo {author} {\bibfnamefont {J.}~\bibnamefont
  {Stachel}}, \bibinfo {author} {\bibfnamefont {A.}~\bibnamefont {Andronic}},
  \bibinfo {author} {\bibfnamefont {P.}~\bibnamefont {Braun-Munzinger}}, \ and\
  \bibinfo {author} {\bibfnamefont {K.}~\bibnamefont {Redlich}},\ }\bibfield
  {booktitle} {\emph {\bibinfo {booktitle} {{Proceedings, 14th International
  Conference on Strangeness in Quark Matter (SQM 2013): Birmingham, UK, July
  22-27, 2013}}},\ }\href {\doibase 10.1088/1742-6596/509/1/012019} {\bibfield
  {journal} {\bibinfo  {journal} {J. Phys. Conf. Ser.}\ }\textbf {\bibinfo
  {volume} {509}},\ \bibinfo {pages} {012019} (\bibinfo {year} {2014})},\
  \Eprint {http://arxiv.org/abs/1311.4662} {arXiv:1311.4662 [nucl-th]}
  \BibitemShut {NoStop}%
\bibitem [{\citenamefont {Floris}(2014)}]{Floris:2014pta}%
  \BibitemOpen
  \bibfield  {author} {\bibinfo {author} {\bibfnamefont {M.}~\bibnamefont
  {Floris}},\ }\bibfield  {booktitle} {\emph {\bibinfo {booktitle}
  {{Proceedings, 24th International Conference on Ultra-Relativistic
  Nucleus-Nucleus Collisions (Quark Matter 2014): Darmstadt, Germany, May
  19-24, 2014}}},\ }\href {\doibase 10.1016/j.nuclphysa.2014.09.002} {\bibfield
   {journal} {\bibinfo  {journal} {Nucl. Phys.}\ }\textbf {\bibinfo {volume}
  {A931}},\ \bibinfo {pages} {103} (\bibinfo {year} {2014})},\ \Eprint
  {http://arxiv.org/abs/1408.6403} {arXiv:1408.6403 [nucl-ex]} \BibitemShut
  {NoStop}%
\bibitem [{\citenamefont {Adam}\ \emph {et~al.}(2017)\citenamefont {Adam} \emph
  {et~al.}}]{ALICE:2017jyt}%
  \BibitemOpen
  \bibfield  {author} {\bibinfo {author} {\bibfnamefont {J.}~\bibnamefont
  {Adam}} \emph {et~al.} (\bibinfo {collaboration} {ALICE}),\ }\href {\doibase
  10.1038/nphys4111} {\bibfield  {journal} {\bibinfo  {journal} {Nature Phys.}\
  }\textbf {\bibinfo {volume} {13}},\ \bibinfo {pages} {535} (\bibinfo {year}
  {2017})},\ \Eprint {http://arxiv.org/abs/1606.07424} {arXiv:1606.07424
  [nucl-ex]} \BibitemShut {NoStop}%
\bibitem [{\citenamefont {Rafelski}\ and\ \citenamefont
  {Danos}(1980)}]{Rafelski:1980gk}%
  \BibitemOpen
  \bibfield  {author} {\bibinfo {author} {\bibfnamefont {J.}~\bibnamefont
  {Rafelski}}\ and\ \bibinfo {author} {\bibfnamefont {M.}~\bibnamefont
  {Danos}},\ }\href {\doibase 10.1016/0370-2693(80)90601-2} {\bibfield
  {journal} {\bibinfo  {journal} {Phys. Lett.}\ }\textbf {\bibinfo {volume}
  {97B}},\ \bibinfo {pages} {279} (\bibinfo {year} {1980})}\BibitemShut
  {NoStop}%
\bibitem [{\citenamefont {Hagedorn}\ and\ \citenamefont
  {Redlich}(1985)}]{Hagedorn:1984uy}%
  \BibitemOpen
  \bibfield  {author} {\bibinfo {author} {\bibfnamefont {R.}~\bibnamefont
  {Hagedorn}}\ and\ \bibinfo {author} {\bibfnamefont {K.}~\bibnamefont
  {Redlich}},\ }\href {\doibase 10.1007/BF01436508} {\bibfield  {journal}
  {\bibinfo  {journal} {Z. Phys.}\ }\textbf {\bibinfo {volume} {C27}},\
  \bibinfo {pages} {541} (\bibinfo {year} {1985})}\BibitemShut {NoStop}%
\bibitem [{\citenamefont {Hamieh}\ \emph {et~al.}(2000)\citenamefont {Hamieh},
  \citenamefont {Redlich},\ and\ \citenamefont {Tounsi}}]{Hamieh:2000tk}%
  \BibitemOpen
  \bibfield  {author} {\bibinfo {author} {\bibfnamefont {S.}~\bibnamefont
  {Hamieh}}, \bibinfo {author} {\bibfnamefont {K.}~\bibnamefont {Redlich}}, \
  and\ \bibinfo {author} {\bibfnamefont {A.}~\bibnamefont {Tounsi}},\ }\href
  {\doibase 10.1016/S0370-2693(00)00762-0} {\bibfield  {journal} {\bibinfo
  {journal} {Phys. Lett.}\ }\textbf {\bibinfo {volume} {B486}},\ \bibinfo
  {pages} {61} (\bibinfo {year} {2000})},\ \Eprint
  {http://arxiv.org/abs/hep-ph/0006024} {arXiv:hep-ph/0006024 [hep-ph]}
  \BibitemShut {NoStop}%
\bibitem [{\citenamefont {Vovchenko}\ \emph
  {et~al.}(2018{\natexlab{a}})\citenamefont {Vovchenko}, \citenamefont
  {Dönigus},\ and\ \citenamefont {Stoecker}}]{Vovchenko:2018fiy}%
  \BibitemOpen
  \bibfield  {author} {\bibinfo {author} {\bibfnamefont {V.}~\bibnamefont
  {Vovchenko}}, \bibinfo {author} {\bibfnamefont {B.}~\bibnamefont {Dönigus}},
  \ and\ \bibinfo {author} {\bibfnamefont {H.}~\bibnamefont {Stoecker}},\
  }\href {\doibase 10.1016/j.physletb.2018.08.041} {\bibfield  {journal}
  {\bibinfo  {journal} {Phys. Lett.}\ }\textbf {\bibinfo {volume} {B785}},\
  \bibinfo {pages} {171} (\bibinfo {year} {2018}{\natexlab{a}})},\ \Eprint
  {http://arxiv.org/abs/1808.05245} {arXiv:1808.05245 [hep-ph]} \BibitemShut
  {NoStop}%
\bibitem [{\citenamefont {Becattini}(1996)}]{Becattini:1995if}%
  \BibitemOpen
  \bibfield  {author} {\bibinfo {author} {\bibfnamefont {F.}~\bibnamefont
  {Becattini}},\ }\href {\doibase 10.1007/BF02907431} {\bibfield  {journal}
  {\bibinfo  {journal} {Z. Phys.}\ }\textbf {\bibinfo {volume} {C69}},\
  \bibinfo {pages} {485} (\bibinfo {year} {1996})}\BibitemShut {NoStop}%
\bibitem [{\citenamefont {Becattini}\ and\ \citenamefont
  {Heinz}(1997)}]{Becattini:1997rv}%
  \BibitemOpen
  \bibfield  {author} {\bibinfo {author} {\bibfnamefont {F.}~\bibnamefont
  {Becattini}}\ and\ \bibinfo {author} {\bibfnamefont {U.~W.}\ \bibnamefont
  {Heinz}},\ }\href {\doibase 10.1007/s002880050551} {\bibfield  {journal}
  {\bibinfo  {journal} {Z. Phys.}\ }\textbf {\bibinfo {volume} {C76}},\
  \bibinfo {pages} {269} (\bibinfo {year} {1997})},\ \bibinfo {note} {[Erratum:
  Z. Phys.C76,578(1997)]},\ \Eprint {http://arxiv.org/abs/hep-ph/9702274}
  {arXiv:hep-ph/9702274 [hep-ph]} \BibitemShut {NoStop}%
\bibitem [{\citenamefont {Vovchenko}\ \emph
  {et~al.}(2018{\natexlab{b}})\citenamefont {Vovchenko}, \citenamefont
  {Gorenstein},\ and\ \citenamefont {Stoecker}}]{Vovchenko:2018fmh}%
  \BibitemOpen
  \bibfield  {author} {\bibinfo {author} {\bibfnamefont {V.}~\bibnamefont
  {Vovchenko}}, \bibinfo {author} {\bibfnamefont {M.~I.}\ \bibnamefont
  {Gorenstein}}, \ and\ \bibinfo {author} {\bibfnamefont {H.}~\bibnamefont
  {Stoecker}},\ }\href {\doibase 10.1103/PhysRevC.98.034906} {\bibfield
  {journal} {\bibinfo  {journal} {Phys. Rev.}\ }\textbf {\bibinfo {volume}
  {C98}},\ \bibinfo {pages} {034906} (\bibinfo {year} {2018}{\natexlab{b}})},\
  \Eprint {http://arxiv.org/abs/1807.02079} {arXiv:1807.02079 [nucl-th]}
  \BibitemShut {NoStop}%
\bibitem [{\citenamefont {Olive}\ \emph {et~al.}(2014)\citenamefont {Olive}
  \emph {et~al.}}]{Agashe:2014kda}%
  \BibitemOpen
  \bibfield  {author} {\bibinfo {author} {\bibfnamefont {K.~A.}\ \bibnamefont
  {Olive}} \emph {et~al.} (\bibinfo {collaboration} {Particle Data Group}),\
  }\href {\doibase 10.1088/1674-1137/38/9/090001} {\bibfield  {journal}
  {\bibinfo  {journal} {Chin. Phys.}\ }\textbf {\bibinfo {volume} {C38}},\
  \bibinfo {pages} {090001} (\bibinfo {year} {2014})}\BibitemShut {NoStop}%
\bibitem [{\citenamefont {Sharma}\ \emph {et~al.}(2019)\citenamefont {Sharma},
  \citenamefont {Cleymans}, \citenamefont {Hippolyte},\ and\ \citenamefont
  {Paradza}}]{Sharma:2018jqf}%
  \BibitemOpen
  \bibfield  {author} {\bibinfo {author} {\bibfnamefont {N.}~\bibnamefont
  {Sharma}}, \bibinfo {author} {\bibfnamefont {J.}~\bibnamefont {Cleymans}},
  \bibinfo {author} {\bibfnamefont {B.}~\bibnamefont {Hippolyte}}, \ and\
  \bibinfo {author} {\bibfnamefont {M.}~\bibnamefont {Paradza}},\ }\href
  {\doibase 10.1103/PhysRevC.99.044914} {\bibfield  {journal} {\bibinfo
  {journal} {Phys. Rev.}\ }\textbf {\bibinfo {volume} {C99}},\ \bibinfo {pages}
  {044914} (\bibinfo {year} {2019})},\ \Eprint
  {http://arxiv.org/abs/1811.00399} {arXiv:1811.00399 [hep-ph]} \BibitemShut
  {NoStop}%
\bibitem [{\citenamefont {Vovchenko}\ and\ \citenamefont
  {Stoecker}(2019)}]{Vovchenko:2019pjl}%
  \BibitemOpen
  \bibfield  {author} {\bibinfo {author} {\bibfnamefont {V.}~\bibnamefont
  {Vovchenko}}\ and\ \bibinfo {author} {\bibfnamefont {H.}~\bibnamefont
  {Stoecker}},\ }\href {\doibase 10.1016/j.cpc.2019.06.024} {\bibfield
  {journal} {\bibinfo  {journal} {Comput. Phys. Commun.}\ }\textbf {\bibinfo
  {volume} {244}},\ \bibinfo {pages} {295} (\bibinfo {year} {2019})},\ \Eprint
  {http://arxiv.org/abs/1901.05249} {arXiv:1901.05249 [nucl-th]} \BibitemShut
  {NoStop}%
\bibitem [{\citenamefont {Vovchenko}(2019)}]{CSM-github}%
  \BibitemOpen
  \bibfield  {author} {\bibinfo {author} {\bibfnamefont {V.}~\bibnamefont
  {Vovchenko}},\ }\href {https://github.com/vlvovch/CSM} {\enquote {\bibinfo
  {title} {Canonical statistical model applications},}\ }\bibinfo
  {howpublished} {GitHub repository} (\bibinfo {year} {2019}),\ \bibinfo {note}
  {[Online; accessed 10-September-2019]}\BibitemShut {NoStop}%
\bibitem [{\citenamefont {Begun}\ \emph {et~al.}(2004)\citenamefont {Begun},
  \citenamefont {Gazdzicki}, \citenamefont {Gorenstein},\ and\ \citenamefont
  {Zozulya}}]{Begun:2004gs}%
  \BibitemOpen
  \bibfield  {author} {\bibinfo {author} {\bibfnamefont {V.~V.}\ \bibnamefont
  {Begun}}, \bibinfo {author} {\bibfnamefont {M.}~\bibnamefont {Gazdzicki}},
  \bibinfo {author} {\bibfnamefont {M.~I.}\ \bibnamefont {Gorenstein}}, \ and\
  \bibinfo {author} {\bibfnamefont {O.~S.}\ \bibnamefont {Zozulya}},\ }\href
  {\doibase 10.1103/PhysRevC.70.034901} {\bibfield  {journal} {\bibinfo
  {journal} {Phys. Rev.}\ }\textbf {\bibinfo {volume} {C70}},\ \bibinfo {pages}
  {034901} (\bibinfo {year} {2004})},\ \Eprint
  {http://arxiv.org/abs/nucl-th/0404056} {arXiv:nucl-th/0404056 [nucl-th]}
  \BibitemShut {NoStop}%
\bibitem [{\citenamefont {Cleymans}\ \emph {et~al.}(1991)\citenamefont
  {Cleymans}, \citenamefont {Redlich},\ and\ \citenamefont
  {Suhonen}}]{Cleymans:1990mn}%
  \BibitemOpen
  \bibfield  {author} {\bibinfo {author} {\bibfnamefont {J.}~\bibnamefont
  {Cleymans}}, \bibinfo {author} {\bibfnamefont {K.}~\bibnamefont {Redlich}}, \
  and\ \bibinfo {author} {\bibfnamefont {E.}~\bibnamefont {Suhonen}},\ }\href
  {\doibase 10.1007/BF01579571} {\bibfield  {journal} {\bibinfo  {journal} {Z.
  Phys.}\ }\textbf {\bibinfo {volume} {C51}},\ \bibinfo {pages} {137} (\bibinfo
  {year} {1991})}\BibitemShut {NoStop}%
\bibitem [{\citenamefont {Cleymans}\ \emph {et~al.}(1998)\citenamefont
  {Cleymans}, \citenamefont {Elliott}, \citenamefont {Keranen},\ and\
  \citenamefont {Suhonen}}]{Cleymans:1997sw}%
  \BibitemOpen
  \bibfield  {author} {\bibinfo {author} {\bibfnamefont {J.}~\bibnamefont
  {Cleymans}}, \bibinfo {author} {\bibfnamefont {D.}~\bibnamefont {Elliott}},
  \bibinfo {author} {\bibfnamefont {A.}~\bibnamefont {Keranen}}, \ and\
  \bibinfo {author} {\bibfnamefont {E.}~\bibnamefont {Suhonen}},\ }\href
  {\doibase 10.1103/PhysRevC.57.3319} {\bibfield  {journal} {\bibinfo
  {journal} {Phys. Rev.}\ }\textbf {\bibinfo {volume} {C57}},\ \bibinfo {pages}
  {3319} (\bibinfo {year} {1998})},\ \Eprint
  {http://arxiv.org/abs/nucl-th/9711066} {arXiv:nucl-th/9711066 [nucl-th]}
  \BibitemShut {NoStop}%
\bibitem [{\citenamefont {Becattini}\ \emph {et~al.}(2001)\citenamefont
  {Becattini}, \citenamefont {Cleymans}, \citenamefont {Keranen}, \citenamefont
  {Suhonen},\ and\ \citenamefont {Redlich}}]{Becattini:2000jw}%
  \BibitemOpen
  \bibfield  {author} {\bibinfo {author} {\bibfnamefont {F.}~\bibnamefont
  {Becattini}}, \bibinfo {author} {\bibfnamefont {J.}~\bibnamefont {Cleymans}},
  \bibinfo {author} {\bibfnamefont {A.}~\bibnamefont {Keranen}}, \bibinfo
  {author} {\bibfnamefont {E.}~\bibnamefont {Suhonen}}, \ and\ \bibinfo
  {author} {\bibfnamefont {K.}~\bibnamefont {Redlich}},\ }\href {\doibase
  10.1103/PhysRevC.64.024901} {\bibfield  {journal} {\bibinfo  {journal} {Phys.
  Rev.}\ }\textbf {\bibinfo {volume} {C64}},\ \bibinfo {pages} {024901}
  (\bibinfo {year} {2001})},\ \Eprint {http://arxiv.org/abs/hep-ph/0002267}
  {arXiv:hep-ph/0002267 [hep-ph]} \BibitemShut {NoStop}%
\bibitem [{\citenamefont {Braun-Munzinger}\ \emph {et~al.}(2002)\citenamefont
  {Braun-Munzinger}, \citenamefont {Cleymans}, \citenamefont {Oeschler},\ and\
  \citenamefont {Redlich}}]{BraunMunzinger:2001as}%
  \BibitemOpen
  \bibfield  {author} {\bibinfo {author} {\bibfnamefont {P.}~\bibnamefont
  {Braun-Munzinger}}, \bibinfo {author} {\bibfnamefont {J.}~\bibnamefont
  {Cleymans}}, \bibinfo {author} {\bibfnamefont {H.}~\bibnamefont {Oeschler}},
  \ and\ \bibinfo {author} {\bibfnamefont {K.}~\bibnamefont {Redlich}},\ }\href
  {\doibase 10.1016/S0375-9474(01)01257-X} {\bibfield  {journal} {\bibinfo
  {journal} {Nucl. Phys.}\ }\textbf {\bibinfo {volume} {A697}},\ \bibinfo
  {pages} {902} (\bibinfo {year} {2002})},\ \Eprint
  {http://arxiv.org/abs/hep-ph/0106066} {arXiv:hep-ph/0106066 [hep-ph]}
  \BibitemShut {NoStop}%
\bibitem [{\citenamefont {Adamczyk}\ \emph {et~al.}(2017)\citenamefont
  {Adamczyk} \emph {et~al.}}]{Adamczyk:2017iwn}%
  \BibitemOpen
  \bibfield  {author} {\bibinfo {author} {\bibfnamefont {L.}~\bibnamefont
  {Adamczyk}} \emph {et~al.} (\bibinfo {collaboration} {STAR}),\ }\href
  {\doibase 10.1103/PhysRevC.96.044904} {\bibfield  {journal} {\bibinfo
  {journal} {Phys. Rev.}\ }\textbf {\bibinfo {volume} {C96}},\ \bibinfo {pages}
  {044904} (\bibinfo {year} {2017})},\ \Eprint
  {http://arxiv.org/abs/1701.07065} {arXiv:1701.07065 [nucl-ex]} \BibitemShut
  {NoStop}%
\bibitem [{\citenamefont {Vovchenko}\ \emph {et~al.}(2016)\citenamefont
  {Vovchenko}, \citenamefont {Begun},\ and\ \citenamefont
  {Gorenstein}}]{Vovchenko:2015idt}%
  \BibitemOpen
  \bibfield  {author} {\bibinfo {author} {\bibfnamefont {V.}~\bibnamefont
  {Vovchenko}}, \bibinfo {author} {\bibfnamefont {V.~V.}\ \bibnamefont
  {Begun}}, \ and\ \bibinfo {author} {\bibfnamefont {M.~I.}\ \bibnamefont
  {Gorenstein}},\ }\href {\doibase 10.1103/PhysRevC.93.064906} {\bibfield
  {journal} {\bibinfo  {journal} {Phys. Rev.}\ }\textbf {\bibinfo {volume}
  {C93}},\ \bibinfo {pages} {064906} (\bibinfo {year} {2016})},\ \Eprint
  {http://arxiv.org/abs/1512.08025} {arXiv:1512.08025 [nucl-th]} \BibitemShut
  {NoStop}%
\bibitem [{\citenamefont {Cleymans}\ \emph {et~al.}(2006)\citenamefont
  {Cleymans}, \citenamefont {Oeschler}, \citenamefont {Redlich},\ and\
  \citenamefont {Wheaton}}]{Cleymans:2005xv}%
  \BibitemOpen
  \bibfield  {author} {\bibinfo {author} {\bibfnamefont {J.}~\bibnamefont
  {Cleymans}}, \bibinfo {author} {\bibfnamefont {H.}~\bibnamefont {Oeschler}},
  \bibinfo {author} {\bibfnamefont {K.}~\bibnamefont {Redlich}}, \ and\
  \bibinfo {author} {\bibfnamefont {S.}~\bibnamefont {Wheaton}},\ }\href
  {\doibase 10.1103/PhysRevC.73.034905} {\bibfield  {journal} {\bibinfo
  {journal} {Phys. Rev.}\ }\textbf {\bibinfo {volume} {C73}},\ \bibinfo {pages}
  {034905} (\bibinfo {year} {2006})},\ \Eprint
  {http://arxiv.org/abs/hep-ph/0511094} {arXiv:hep-ph/0511094 [hep-ph]}
  \BibitemShut {NoStop}%
\bibitem [{\citenamefont {Castorina}\ and\ \citenamefont
  {Satz}(2014)}]{Castorina:2013mba}%
  \BibitemOpen
  \bibfield  {author} {\bibinfo {author} {\bibfnamefont {P.}~\bibnamefont
  {Castorina}}\ and\ \bibinfo {author} {\bibfnamefont {H.}~\bibnamefont
  {Satz}},\ }\href {\doibase 10.1142/S0218301314500190} {\bibfield  {journal}
  {\bibinfo  {journal} {Int. J. Mod. Phys.}\ }\textbf {\bibinfo {volume}
  {E23}},\ \bibinfo {pages} {1450019} (\bibinfo {year} {2014})},\ \Eprint
  {http://arxiv.org/abs/1310.6932} {arXiv:1310.6932 [hep-ph]} \BibitemShut
  {NoStop}%
\bibitem [{\citenamefont {Jeon}\ and\ \citenamefont
  {Koch}(2000)}]{Jeon:2000wg}%
  \BibitemOpen
  \bibfield  {author} {\bibinfo {author} {\bibfnamefont {S.}~\bibnamefont
  {Jeon}}\ and\ \bibinfo {author} {\bibfnamefont {V.}~\bibnamefont {Koch}},\
  }\href {\doibase 10.1103/PhysRevLett.85.2076} {\bibfield  {journal} {\bibinfo
   {journal} {Phys. Rev. Lett.}\ }\textbf {\bibinfo {volume} {85}},\ \bibinfo
  {pages} {2076} (\bibinfo {year} {2000})},\ \Eprint
  {http://arxiv.org/abs/hep-ph/0003168} {arXiv:hep-ph/0003168 [hep-ph]}
  \BibitemShut {NoStop}%
\bibitem [{\citenamefont {Bzdak}\ \emph {et~al.}(2013)\citenamefont {Bzdak},
  \citenamefont {Koch},\ and\ \citenamefont {Skokov}}]{Bzdak:2012an}%
  \BibitemOpen
  \bibfield  {author} {\bibinfo {author} {\bibfnamefont {A.}~\bibnamefont
  {Bzdak}}, \bibinfo {author} {\bibfnamefont {V.}~\bibnamefont {Koch}}, \ and\
  \bibinfo {author} {\bibfnamefont {V.}~\bibnamefont {Skokov}},\ }\href
  {\doibase 10.1103/PhysRevC.87.014901} {\bibfield  {journal} {\bibinfo
  {journal} {Phys. Rev.}\ }\textbf {\bibinfo {volume} {C87}},\ \bibinfo {pages}
  {014901} (\bibinfo {year} {2013})},\ \Eprint {http://arxiv.org/abs/1203.4529}
  {arXiv:1203.4529 [hep-ph]} \BibitemShut {NoStop}%
\bibitem [{\citenamefont {Braun-Munzinger}\ \emph {et~al.}(2017)\citenamefont
  {Braun-Munzinger}, \citenamefont {Rustamov},\ and\ \citenamefont
  {Stachel}}]{Braun-Munzinger:2016yjz}%
  \BibitemOpen
  \bibfield  {author} {\bibinfo {author} {\bibfnamefont {P.}~\bibnamefont
  {Braun-Munzinger}}, \bibinfo {author} {\bibfnamefont {A.}~\bibnamefont
  {Rustamov}}, \ and\ \bibinfo {author} {\bibfnamefont {J.}~\bibnamefont
  {Stachel}},\ }\href {\doibase 10.1016/j.nuclphysa.2017.01.011} {\bibfield
  {journal} {\bibinfo  {journal} {Nucl. Phys.}\ }\textbf {\bibinfo {volume}
  {A960}},\ \bibinfo {pages} {114} (\bibinfo {year} {2017})},\ \Eprint
  {http://arxiv.org/abs/1612.00702} {arXiv:1612.00702 [nucl-th]} \BibitemShut
  {NoStop}%
\bibitem [{\citenamefont {Pruneau}(2019)}]{Pruneau:2019baa}%
  \BibitemOpen
  \bibfield  {author} {\bibinfo {author} {\bibfnamefont {C.~A.}\ \bibnamefont
  {Pruneau}},\ }\href {\doibase 10.1103/PhysRevC.100.034905} {\bibfield
  {journal} {\bibinfo  {journal} {Phys. Rev.}\ }\textbf {\bibinfo {volume}
  {C100}},\ \bibinfo {pages} {034905} (\bibinfo {year} {2019})},\ \Eprint
  {http://arxiv.org/abs/1903.04591} {arXiv:1903.04591 [nucl-th]} \BibitemShut
  {NoStop}%
\bibitem [{\citenamefont {Rustamov}(2017)}]{Rustamov:2017lio}%
  \BibitemOpen
  \bibfield  {author} {\bibinfo {author} {\bibfnamefont {A.}~\bibnamefont
  {Rustamov}} (\bibinfo {collaboration} {ALICE}),\ }\bibfield  {booktitle}
  {\emph {\bibinfo {booktitle} {{Proceedings, 26th International Conference on
  Ultra-relativistic Nucleus-Nucleus Collisions (Quark Matter 2017): Chicago,
  Illinois, USA, February 5-11, 2017}}},\ }\href {\doibase
  10.1016/j.nuclphysa.2017.05.111} {\bibfield  {journal} {\bibinfo  {journal}
  {Nucl. Phys.}\ }\textbf {\bibinfo {volume} {A967}},\ \bibinfo {pages} {453}
  (\bibinfo {year} {2017})},\ \Eprint {http://arxiv.org/abs/1704.05329}
  {arXiv:1704.05329 [nucl-ex]} \BibitemShut {NoStop}%
\bibitem [{\citenamefont {Bass}\ \emph {et~al.}(2000)\citenamefont {Bass},
  \citenamefont {Danielewicz},\ and\ \citenamefont {Pratt}}]{Bass:2000az}%
  \BibitemOpen
  \bibfield  {author} {\bibinfo {author} {\bibfnamefont {S.~A.}\ \bibnamefont
  {Bass}}, \bibinfo {author} {\bibfnamefont {P.}~\bibnamefont {Danielewicz}}, \
  and\ \bibinfo {author} {\bibfnamefont {S.}~\bibnamefont {Pratt}},\ }\href
  {\doibase 10.1103/PhysRevLett.85.2689} {\bibfield  {journal} {\bibinfo
  {journal} {Phys. Rev. Lett.}\ }\textbf {\bibinfo {volume} {85}},\ \bibinfo
  {pages} {2689} (\bibinfo {year} {2000})},\ \Eprint
  {http://arxiv.org/abs/nucl-th/0005044} {arXiv:nucl-th/0005044 [nucl-th]}
  \BibitemShut {NoStop}%
\bibitem [{\citenamefont {Pratt}(2012)}]{Pratt:2011bc}%
  \BibitemOpen
  \bibfield  {author} {\bibinfo {author} {\bibfnamefont {S.}~\bibnamefont
  {Pratt}},\ }\href {\doibase 10.1103/PhysRevC.85.014904} {\bibfield  {journal}
  {\bibinfo  {journal} {Phys. Rev.}\ }\textbf {\bibinfo {volume} {C85}},\
  \bibinfo {pages} {014904} (\bibinfo {year} {2012})},\ \Eprint
  {http://arxiv.org/abs/1109.3647} {arXiv:1109.3647 [nucl-th]} \BibitemShut
  {NoStop}%
\bibitem [{\citenamefont {Shuryak}\ and\ \citenamefont
  {Zahed}(2013)}]{Shuryak:2013ke}%
  \BibitemOpen
  \bibfield  {author} {\bibinfo {author} {\bibfnamefont {E.}~\bibnamefont
  {Shuryak}}\ and\ \bibinfo {author} {\bibfnamefont {I.}~\bibnamefont
  {Zahed}},\ }\href {\doibase 10.1103/PhysRevC.88.044915} {\bibfield  {journal}
  {\bibinfo  {journal} {Phys. Rev.}\ }\textbf {\bibinfo {volume} {C88}},\
  \bibinfo {pages} {044915} (\bibinfo {year} {2013})},\ \Eprint
  {http://arxiv.org/abs/1301.4470} {arXiv:1301.4470 [hep-ph]} \BibitemShut
  {NoStop}%
\bibitem [{\citenamefont {Bellwied}\ \emph {et~al.}(2013)\citenamefont
  {Bellwied}, \citenamefont {Borsanyi}, \citenamefont {Fodor}, \citenamefont
  {Katz},\ and\ \citenamefont {Ratti}}]{Bellwied:2013cta}%
  \BibitemOpen
  \bibfield  {author} {\bibinfo {author} {\bibfnamefont {R.}~\bibnamefont
  {Bellwied}}, \bibinfo {author} {\bibfnamefont {S.}~\bibnamefont {Borsanyi}},
  \bibinfo {author} {\bibfnamefont {Z.}~\bibnamefont {Fodor}}, \bibinfo
  {author} {\bibfnamefont {S.~D.}\ \bibnamefont {Katz}}, \ and\ \bibinfo
  {author} {\bibfnamefont {C.}~\bibnamefont {Ratti}},\ }\href {\doibase
  10.1103/PhysRevLett.111.202302} {\bibfield  {journal} {\bibinfo  {journal}
  {Phys. Rev. Lett.}\ }\textbf {\bibinfo {volume} {111}},\ \bibinfo {pages}
  {202302} (\bibinfo {year} {2013})},\ \Eprint {http://arxiv.org/abs/1305.6297}
  {arXiv:1305.6297 [hep-lat]} \BibitemShut {NoStop}%
\bibitem [{\citenamefont {Chatterjee}\ \emph {et~al.}(2017)\citenamefont
  {Chatterjee}, \citenamefont {Dash},\ and\ \citenamefont
  {Mohanty}}]{Chatterjee:2016cog}%
  \BibitemOpen
  \bibfield  {author} {\bibinfo {author} {\bibfnamefont {S.}~\bibnamefont
  {Chatterjee}}, \bibinfo {author} {\bibfnamefont {A.~K.}\ \bibnamefont
  {Dash}}, \ and\ \bibinfo {author} {\bibfnamefont {B.}~\bibnamefont
  {Mohanty}},\ }\href {\doibase 10.1088/1361-6471/aa8857} {\bibfield  {journal}
  {\bibinfo  {journal} {J. Phys.}\ }\textbf {\bibinfo {volume} {G44}},\
  \bibinfo {pages} {105106} (\bibinfo {year} {2017})},\ \Eprint
  {http://arxiv.org/abs/1608.00643} {arXiv:1608.00643 [nucl-th]} \BibitemShut
  {NoStop}%
\bibitem [{\citenamefont {Koch}\ \emph {et~al.}(1986)\citenamefont {Koch},
  \citenamefont {Muller},\ and\ \citenamefont {Rafelski}}]{Koch:1986ud}%
  \BibitemOpen
  \bibfield  {author} {\bibinfo {author} {\bibfnamefont {P.}~\bibnamefont
  {Koch}}, \bibinfo {author} {\bibfnamefont {B.}~\bibnamefont {Muller}}, \ and\
  \bibinfo {author} {\bibfnamefont {J.}~\bibnamefont {Rafelski}},\ }\href
  {\doibase 10.1016/0370-1573(86)90096-7} {\bibfield  {journal} {\bibinfo
  {journal} {Phys. Rept.}\ }\textbf {\bibinfo {volume} {142}},\ \bibinfo
  {pages} {167} (\bibinfo {year} {1986})}\BibitemShut {NoStop}%
\bibitem [{\citenamefont {Rafelski}(1991)}]{Rafelski:1991rh}%
  \BibitemOpen
  \bibfield  {author} {\bibinfo {author} {\bibfnamefont {J.}~\bibnamefont
  {Rafelski}},\ }\href {\doibase 10.1016/0370-2693(91)91576-H} {\bibfield
  {journal} {\bibinfo  {journal} {Phys. Lett.}\ }\textbf {\bibinfo {volume}
  {B262}},\ \bibinfo {pages} {333} (\bibinfo {year} {1991})}\BibitemShut
  {NoStop}%
\bibitem [{\citenamefont {Abelev}\ \emph
  {et~al.}(2015{\natexlab{b}})\citenamefont {Abelev} \emph
  {et~al.}}]{Abelev:2014qqa}%
  \BibitemOpen
  \bibfield  {author} {\bibinfo {author} {\bibfnamefont {B.~B.}\ \bibnamefont
  {Abelev}} \emph {et~al.} (\bibinfo {collaboration} {ALICE}),\ }\href
  {\doibase 10.1140/epjc/s10052-014-3191-x} {\bibfield  {journal} {\bibinfo
  {journal} {Eur. Phys. J.}\ }\textbf {\bibinfo {volume} {C75}},\ \bibinfo
  {pages} {1} (\bibinfo {year} {2015}{\natexlab{b}})},\ \Eprint
  {http://arxiv.org/abs/1406.3206} {arXiv:1406.3206 [nucl-ex]} \BibitemShut
  {NoStop}%
\bibitem [{\citenamefont {Acharya}\ \emph
  {et~al.}(2019{\natexlab{b}})\citenamefont {Acharya} \emph
  {et~al.}}]{Acharya:2018qnp}%
  \BibitemOpen
  \bibfield  {author} {\bibinfo {author} {\bibfnamefont {S.}~\bibnamefont
  {Acharya}} \emph {et~al.} (\bibinfo {collaboration} {ALICE}),\ }\href
  {\doibase 10.1103/PhysRevC.99.064901} {\bibfield  {journal} {\bibinfo
  {journal} {Phys. Rev.}\ }\textbf {\bibinfo {volume} {C99}},\ \bibinfo {pages}
  {064901} (\bibinfo {year} {2019}{\natexlab{b}})},\ \Eprint
  {http://arxiv.org/abs/1805.04365} {arXiv:1805.04365 [nucl-ex]} \BibitemShut
  {NoStop}%
\bibitem [{\citenamefont {Adamova}\ \emph {et~al.}(2017)\citenamefont {Adamova}
  \emph {et~al.}}]{Adamova:2017elh}%
  \BibitemOpen
  \bibfield  {author} {\bibinfo {author} {\bibfnamefont {D.}~\bibnamefont
  {Adamova}} \emph {et~al.} (\bibinfo {collaboration} {ALICE}),\ }\href
  {\doibase 10.1140/epjc/s10052-017-4943-1} {\bibfield  {journal} {\bibinfo
  {journal} {Eur. Phys. J.}\ }\textbf {\bibinfo {volume} {C77}},\ \bibinfo
  {pages} {389} (\bibinfo {year} {2017})},\ \Eprint
  {http://arxiv.org/abs/1701.07797} {arXiv:1701.07797 [nucl-ex]} \BibitemShut
  {NoStop}%
\bibitem [{\citenamefont {Acharya}\ \emph
  {et~al.}(2019{\natexlab{c}})\citenamefont {Acharya} \emph
  {et~al.}}]{ALICE:2018ewo}%
  \BibitemOpen
  \bibfield  {author} {\bibinfo {author} {\bibfnamefont {S.}~\bibnamefont
  {Acharya}} \emph {et~al.} (\bibinfo {collaboration} {ALICE}),\ }\href
  {\doibase 10.1103/PhysRevC.99.024905} {\bibfield  {journal} {\bibinfo
  {journal} {Phys. Rev.}\ }\textbf {\bibinfo {volume} {C99}},\ \bibinfo {pages}
  {024905} (\bibinfo {year} {2019}{\natexlab{c}})},\ \Eprint
  {http://arxiv.org/abs/1805.04361} {arXiv:1805.04361 [nucl-ex]} \BibitemShut
  {NoStop}%
\bibitem [{\citenamefont {Kalinak}(2017{\natexlab{a}})}]{Kalinak}%
  \BibitemOpen
  \bibfield  {author} {\bibinfo {author} {\bibfnamefont {P.}~\bibnamefont
  {Kalinak}} (\bibinfo {collaboration} {ALICE}),\ }\href
  {https://indico.cern.ch/event/466934/contributions/2588311/attachments/1492435/2320500/StrangenessProduction5TeV_EPS_HEP2017_v07.pdf}
  {\bibfield  {journal} {\bibinfo  {journal} {European Physical Society
  Conference on High Energy Physics (EPS-HEP 2017): Venice, Italy}\ } (\bibinfo
  {year} {2017}{\natexlab{a}})}\BibitemShut {NoStop}%
\bibitem [{\citenamefont {Kalinak}(2017{\natexlab{b}})}]{Kalinak:2017xll}%
  \BibitemOpen
  \bibfield  {author} {\bibinfo {author} {\bibfnamefont {P.}~\bibnamefont
  {Kalinak}} (\bibinfo {collaboration} {ALICE}),\ }\bibfield  {booktitle}
  {\emph {\bibinfo {booktitle} {{Proceedings, 2017 European Physical Society
  Conference on High Energy Physics (EPS-HEP 2017): Venice, Italy, July 5-12,
  2017}}},\ }\href {\doibase 10.22323/1.314.0168} {\bibfield  {journal}
  {\bibinfo  {journal} {PoS}\ }\textbf {\bibinfo {volume} {EPS-HEP2017}},\
  \bibinfo {pages} {168} (\bibinfo {year} {2017}{\natexlab{b}})}\BibitemShut
  {NoStop}%
\bibitem [{\citenamefont {Bebie}\ \emph {et~al.}(1992)\citenamefont {Bebie},
  \citenamefont {Gerber}, \citenamefont {Goity},\ and\ \citenamefont
  {Leutwyler}}]{Bebie:1991ij}%
  \BibitemOpen
  \bibfield  {author} {\bibinfo {author} {\bibfnamefont {H.}~\bibnamefont
  {Bebie}}, \bibinfo {author} {\bibfnamefont {P.}~\bibnamefont {Gerber}},
  \bibinfo {author} {\bibfnamefont {J.~L.}\ \bibnamefont {Goity}}, \ and\
  \bibinfo {author} {\bibfnamefont {H.}~\bibnamefont {Leutwyler}},\ }\href
  {\doibase 10.1016/0550-3213(92)90005-V} {\bibfield  {journal} {\bibinfo
  {journal} {Nucl. Phys.}\ }\textbf {\bibinfo {volume} {B378}},\ \bibinfo
  {pages} {95} (\bibinfo {year} {1992})}\BibitemShut {NoStop}%
\bibitem [{\citenamefont {Vovchenko}\ \emph {et~al.}(2019)\citenamefont
  {Vovchenko}, \citenamefont {Gallmeister}, \citenamefont {Schaffner-Bielich},\
  and\ \citenamefont {Greiner}}]{Vovchenko:2019aoz}%
  \BibitemOpen
  \bibfield  {author} {\bibinfo {author} {\bibfnamefont {V.}~\bibnamefont
  {Vovchenko}}, \bibinfo {author} {\bibfnamefont {K.}~\bibnamefont
  {Gallmeister}}, \bibinfo {author} {\bibfnamefont {J.}~\bibnamefont
  {Schaffner-Bielich}}, \ and\ \bibinfo {author} {\bibfnamefont
  {C.}~\bibnamefont {Greiner}},\ }\href@noop {} {\  (\bibinfo {year} {2019})},\
  \Eprint {http://arxiv.org/abs/1903.10024} {arXiv:1903.10024 [hep-ph]}
  \BibitemShut {NoStop}%
\bibitem [{\citenamefont {Steinheimer}\ \emph {et~al.}(2017)\citenamefont
  {Steinheimer}, \citenamefont {Aichelin}, \citenamefont {Bleicher},\ and\
  \citenamefont {Stöcker}}]{Steinheimer:2017vju}%
  \BibitemOpen
  \bibfield  {author} {\bibinfo {author} {\bibfnamefont {J.}~\bibnamefont
  {Steinheimer}}, \bibinfo {author} {\bibfnamefont {J.}~\bibnamefont
  {Aichelin}}, \bibinfo {author} {\bibfnamefont {M.}~\bibnamefont {Bleicher}},
  \ and\ \bibinfo {author} {\bibfnamefont {H.}~\bibnamefont {Stöcker}},\
  }\href {\doibase 10.1103/PhysRevC.95.064902} {\bibfield  {journal} {\bibinfo
  {journal} {Phys. Rev.}\ }\textbf {\bibinfo {volume} {C95}},\ \bibinfo {pages}
  {064902} (\bibinfo {year} {2017})},\ \Eprint
  {http://arxiv.org/abs/1703.06638} {arXiv:1703.06638 [nucl-th]} \BibitemShut
  {NoStop}%
\bibitem [{\citenamefont {Bazavov}\ \emph {et~al.}(2019)\citenamefont {Bazavov}
  \emph {et~al.}}]{Bazavov:2018mes}%
  \BibitemOpen
  \bibfield  {author} {\bibinfo {author} {\bibfnamefont {A.}~\bibnamefont
  {Bazavov}} \emph {et~al.} (\bibinfo {collaboration} {HotQCD}),\ }\href
  {\doibase 10.1016/j.physletb.2019.05.013} {\bibfield  {journal} {\bibinfo
  {journal} {Phys. Lett.}\ }\textbf {\bibinfo {volume} {B795}},\ \bibinfo
  {pages} {15} (\bibinfo {year} {2019})},\ \Eprint
  {http://arxiv.org/abs/1812.08235} {arXiv:1812.08235 [hep-lat]} \BibitemShut
  {NoStop}%
\end{thebibliography}%


\end{document}